\documentclass[pre,superscriptaddress,letterpaper]{revtex4}

\usepackage{graphicx}
\usepackage{amssymb}
\usepackage{chemarr}
\usepackage{appendix}
\usepackage{setspace}
\def\e{{\bf e}}

\begin{document}

\title{Dynamics and Efficiency of Brownian Rotors}

Category: Physics, Biophysics

\author{Wolfgang R. Bauer\footnote{corresponding author
% \\
%Tel. +49-931-201-36327\\
%Fax  +49-931-201-36291\\
}} \email{w.bauer@medizin.uni-wuerzburg.de} \affiliation{
Medizinische Universit\"atsklinik 1, Josef Schneider Stra{\ss}e 2,
D-97080 W\"urzburg, Germany }
\author{Walter Nadler}
\email{w.nadler@fz-juelich.de} \affiliation{
J\"ulich Supercomputing Centre,  
Forschungszentrum J\"ulich,
D-52425 J\"ulich, Germany
%Computational Nano- and Biophysics Group,
%Department of Physics, Michigan Technological University, 1400
%Townsend Drive, Houghton, MI 49931-1295, USA
%\footnote{present address}
}

\date{\today}

\begin{abstract}
Brownian rotors play an important role in biological systems and 
%perhaps 
in future nano-technological applications. However the mechanisms determining their dynamics, efficiency and performance remain to be characterized. Here the F0 portion of the F-ATP synthase is considered as a paradigm of a Brownian rotor. In a generic analytical  model we analyze the stochastic rotation of F0-like motors as a function of the driving free energy difference and of the free energy profile the rotor is subjected to. The latter is composed of the rotor interaction with its 
%embedding 
surroundings, of the free energy of chemical transitions, and of the workload. The dynamics and mechanical efficiency of the rotor depends on the magnitude of its stochastic motion driven by the free energy energy difference and its rectification on the reaction-diffusion path. We analyze which free energy profiles provide maximum flow and how their arrangement on the underlying reaction-diffusion path affects rectification and -- by this -- the efficiency. 
\end{abstract}

\maketitle

\narrowtext

\section*{Introduction}
Molecular motors play a central role in biological systems where they convert a free energy  difference into mechanical force. There exist two fundamental mechanisms by which this can be achieved \cite{Leigh}. The free energy difference, coming e.g. from a chemical reaction, can be converted directly into mechanical force by some kind of power strike \cite{Wang1998, Wang2000, Oster}. The other mechanism is more sophisticated, since here the mechanical movement is provided by thermal fluctuations of the motor\cite{Astumian2, Peskin, Parrondo}. In the latter case the free energy difference is transformed into entropic forces biasing the direction of theses fluctuations . Both mechanism may also be combined and work synergistically \cite{Bustamante}.  

An archetype of Brownian rotors is the F0 portion of the F-ATP synthase \cite{vik, junge1997, remark}. An electro-chemical
gradient of protons or sodium ions across the inner membrane of mitochondria builds up a free energy difference which is converted into a mechanical torque. The latter drives ATP synthesis, the essential energy carrier in living objects. One accepted descriptive model for the torque generation of the proton driven F-ATP synthase is briefly (Fig.~1): 
The F0 motor consists of a rotary ring, carrying identical hairpin-like protomers in  
which aspartic or glutamic acid residues transport protons along the electro-chemical gradient. The ring is subject to thermal fluctuations and electrostatic interactions with its molecular neighborhood. The latter result from the rotor's interface 
with a positively charged stator, which attracts negatively charged (deprotonated) protomers of the rotor. In addition, hydrophilic/hydrophobic interactions confine these charged protomers to the hydrophilic stator region \cite{vik, junge1997}. The protomer facing the hydrophobic membrane requires release from this constraint through neutralization by a proton coming from the access channels \cite{jungeScience}. These channels have contact to the  respective membrane sides.  The frequency by which the constraint is removed is proportional to the proton concentrations in the channel, or more precisely to their activity when electrical fields are present. A concentration or activity difference of protons between the two channel access sites builds up a corresponding probability difference to find one protonated protomer, i.e. occupied proton carrier. This entropic force directs proton flow towards equilibration of the proton concentration difference. Rectification of this flow by appropriate interactions then produces directed rotation and torque\cite{animation}.

Though there exists valuable and detailed information in literature about the mechanisms of
rotation of the F0 motor \cite{Elston, Xing, Aksimentiev} and its transduction in ATP synthesis \cite{Bockmann} many fundamental questions remain to be solved. How does the free energy profile on the reaction-diffusion path, i.e. ring neighborhood interaction, chemical energy and workload, mediate the effect of the driving free energy difference on proton flow, and how does it rectify this flow to achieve directed rotation? Is there an
optimum arrangement and strength of interactions for maximum rotation speed? In which way does the binding
strength of protons affect the ring rotation? We could recently demonstrate that -- depending on its magnitude -- an
attractive interaction may support or hamper fluctuation driven motion \cite{BauerNadlerpnas}. What is the performance of the ring motor, and when does it reach its maximum?   
Some of the aspects mentioned above were addressed in previous works by simulations or numerical techniques \cite{remark,Elston} in which many details of the biological system are considered. In contrast, we will develop here a generic analytical  model to obtain fundamental relationships that can answer the above mentioned questions. This implies that we consider a model which reduces details of the original biological system but still conserves its fundamental properties. Especially for technical applications this will be of importance for the development of an optimal working (supra) molecular motor / nano-motor which, triggered by chemical breakthroughs \cite{stoddart}, has become an emerging field \cite{Leigh}.

 \begin{figure}
\label{figure1}
\includegraphics[width=11cm, angle=-0]{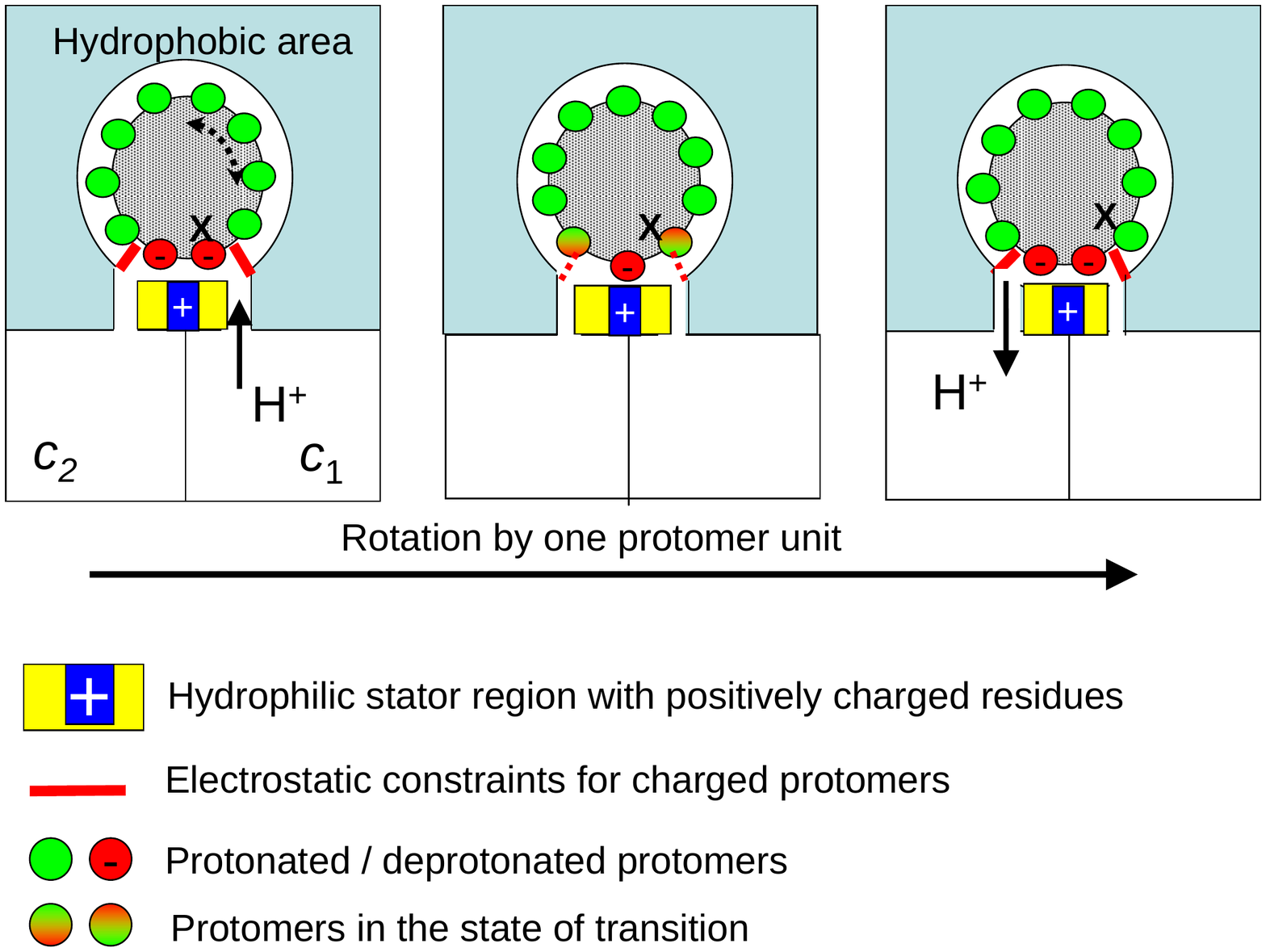}
\caption{Schematic showing the rotation of one protomer unit of the F0 portion of the F-ATP synthase, marked with X. Protons of either membrane side have access to the ring via channels. The concentration gradient $c_1>c_2$ biases counterclockwise rotation (see text), resulting in proton flow in the gradient's direction.}
\end{figure}

The paper is structured as follows: Based on the short qualitative description of the operation mode above, an analytic  model of  the F0 motor is presented. For didactical reasons we start with a full rectified rotor, i.e. a perfect coupling between proton flow and rotation, which is achieved by the assumption that deprotonated protomers are strictly confined to the hydrophilic stator, while protonated protomers are confined to the hydrophobic membrane region \cite{jungeScience}. This implies a simple circular topology of the reaction-diffusion path, on which the optimum arrangement of ring neighborhood interactions providing maximum flow, i.e. rotation speed, is derived. This is done in the absence, and presence of external workload. In the latter case also the performance, defined as external work per time, is analyzed. In the third section we give up the constraint of perfect rectification, which implies a more complex diffusion-reaction path. We analyze the coupling of flow and rotation and the stall workload, determining the efficiency of the rotor as a function of the interactions.

\section*{Motor with Check Valve Mechanism}
\subsection*{No Workload}

\begin{figure}
\label{figure2}
\includegraphics[width=9cm, angle=-0]{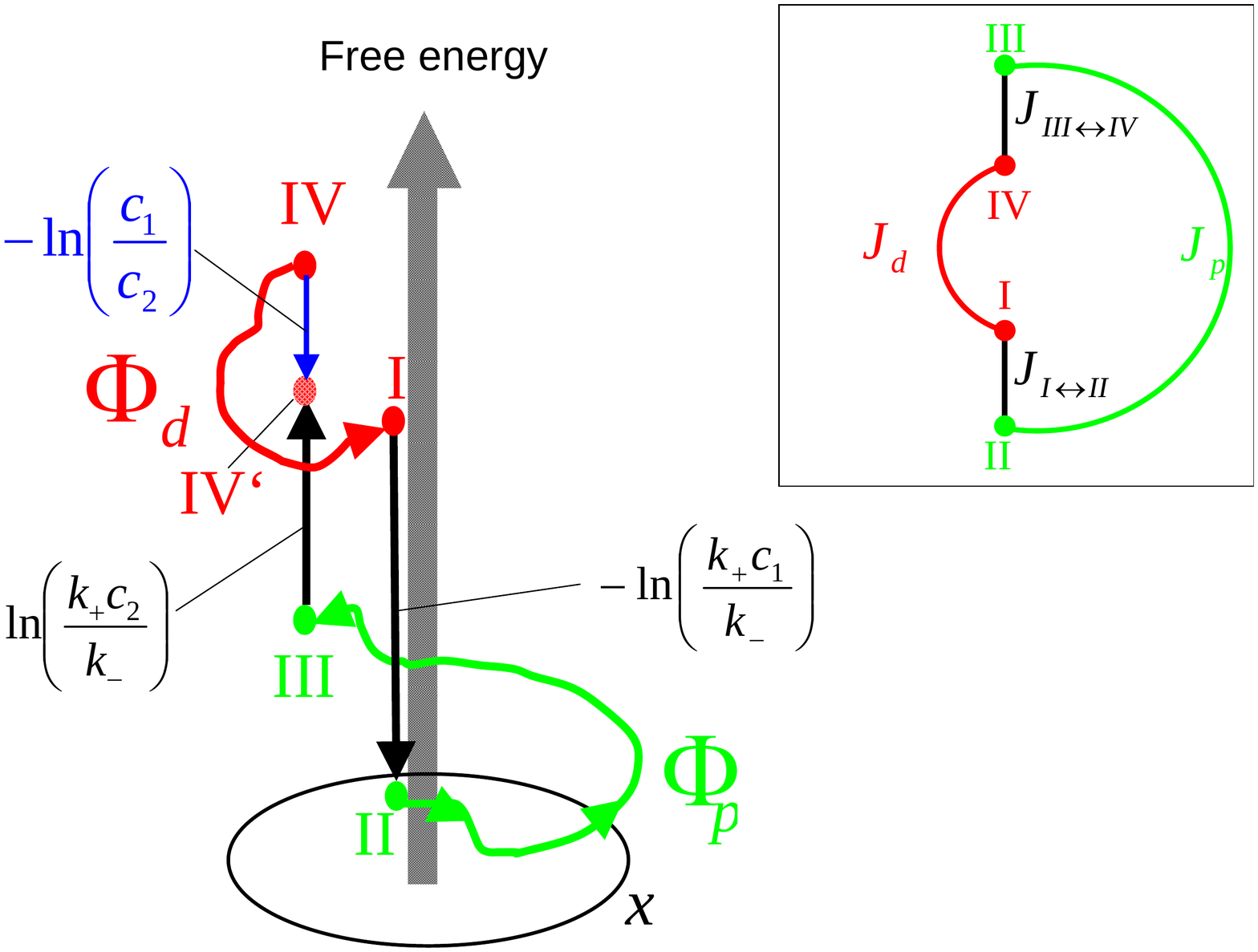}
\caption{Free energy profile of the rotor (deprotonated state red, protonated state green, chemical transitions black) as a function of the cyclic spatial variable $x$. The location of the channels in which protonation/de-protonation takes place is marked by filled circles. Hydrophilic and hydrophobic interactions confine the deprotonated state to the left, the protonated to the right section of the diffusion path (see text), and transitions are solely feasible by chemical reaction. This implies that the diffusion-reaction path has a circular topology (insert). The free energy course during one cycle on this path is marked by arrows ($IV\to I\to II\cdots$). The choice of proton concentrations $c_1>c_2$ makes the rotor run in counterclockwise direction and it returns to point $IV$ (hatched circle) at a lower free energy state (blue) $-\ln( c_1/c_2)$. The insert shows the sections of the diffusion and chemical transition paths forming a circular topology and the corresponding  diffusive flows and chemical fluxes $J_d,\; J_d,\; J_{I\rightleftharpoons II}, J_{III\rightleftharpoons IV}$.}
\end{figure}

In our model we consider rotation by one protomer unit as a cyclic process (Fig.~1-2). Two states are distinguished: a protonated and a deprotonated one, where the latter refers to the maximum number of negatively charged (deprotonated) protomers within the hydrophilic region. Within each state the rotor undergoes diffusive motion within some free energy landscape (Fig.~2), which is determined by the ring neighborhood interactions, e.g. electrostatic forces between deprotonated protomers and the positively charged stator, and hydrophilic/hydrophobic interactions. Transitions between the states occur at the locations of the channels, quantified by reaction rates $k_+,\; k_-$  for protonation and de-protonation respectively,
\begin{equation}
\hbox{deprotonated state}\xrightleftharpoons [k_-]{k_+\;c}\hbox{protonated state}\label{reaction}
\end{equation}
where $c$ is the proton concentration in the channel. As suggested in the original works \cite{vik, junge1997} and stated in Ref.~\cite{jungeScience}, we assume in this section that protomers must be the negatively charged (deprotonated) when facing the positively charged hydrophilic stator region, and neutralized (protonated) when facing the hydrophobic membrane . So in our model very high energy barriers confine the deprotonated and protonated state to two complementary spatial sections which are interconnected by the chemical reaction pathways. This   makes the topology of the diffusion-plus-chemical-reaction pathway that of an oriented circle (Fig.~2), and the free energy difference can drive the motor solely in one direction of rotation. So the high energy barriers act as mechanical check valves rectifying motion and maintaining a perfect coupling of proton flow and rotation.

Formally the state of the rotor is determined by its protonation state $d,\;p$ and its local position within this
state $x$. The dynamics of the system consists of diffusion and chemical transitions which is described by a set of two Smoluchowski equations~\cite{Gardiner} to which chemical reaction terms are added. One obtains for the probability density,
\begin{eqnarray}
\partial_t \rho_{d}(x,t)&=&\partial_x D_d(x)\left[\partial_x-F_d(x)\right]\rho_{d}(x,t) - \Delta\cr\cr
\partial_t \rho_{p}(x,t)&=&\partial_x D_p(x)\left[\partial_x-F_p(x)\right]\rho_{p}(x,t) + \Delta
\;,\label{Smoluchowski}
\end{eqnarray}
where $D_i(x)$ ($i=d,p)$ is the diffusion coefficient, which we assume to be constant in the following, and the force $F_i(x)$ describes the interaction of the rotor with its
surrounding. In the one-dimensional case considered here, this mechanical force can always be derived from a local potential
$F_i(x)=-\Phi_i'(x)$.  The term $\Delta$ derives from the chemical reaction (\ref{reaction}) and describes protonation and de-protonation via
\begin{eqnarray}
\Delta=k_+\; c_1\;\rho_{d}(x,t)\delta(x-x_{I\rightleftharpoons II}) - k_-\;\rho_{p}(x,t)\delta(x-x_{I\rightleftharpoons II})&&+\cr \cr
 k_+\; c_2\;\rho_{d}(x,t)\delta(x-x_{III\rightleftharpoons IV})-k_- \; \rho_{p}(x,t)\delta(x-x_{III\rightleftharpoons IV})&&\;,
\end{eqnarray}
where $c_1,\; c_2$ are the proton concentrations in the respective channels.  The delta distribution $\delta(x-x_{A\rightleftharpoons B})$ locates chemical reactions on the two cross points of chemical transition and respective diffusion paths which are labeled by the roman subscribes $A,B=I,\ldots,IV$ (Fig.~2). Since the rotor must be in some state,  conservation of probability holds, 
\begin{equation}
\int_{x\in S_d} \rho_d(x,t) dx + \int_{x\in S_p} \rho_p(x,t) dx=1 \;,\label{conservationprobability1}
\end{equation}  
where $S_d\; S_d$ are the spatial sections to which the rotor is confined in the deprotonated/ protonated state. 

The stochastic motion of ring rotation is directly related to the diffusive probability flow  which in the respective protonation states has the form
\begin{equation}
j_i(x,t)=-D_i\left[\partial_x-F_i(x)\right]\rho_i(x,t) \; . \label{flow}
\end{equation}
The chemical flux at the transition between deprotonated and protonated state is 
\begin{eqnarray}
j_{I\rightleftharpoons II}(t)&=&k_+\;c_1\;\rho_{d}(x_{I\rightleftharpoons II},t) - k_-\;\rho_{p}(x_{I\rightleftharpoons II},t)\label{flow2}\cr\cr j_{III\rightleftharpoons IV}(t)&=&k_- \; \rho_{p}(x_{III\rightleftharpoons IV},t) -
k_+\;c_2\;\rho_{d}(x_{III\rightleftharpoons IV},t)\label{flow3}\;,
\end{eqnarray}

In the steady state the probability densities become stationary $\rho_i(x,t)\to \rho_i(x), \; i=d,p$ and conservation of flow holds. For diffusive flow this implies $j_i(x,t)\equiv J_i$. In addition the circular topology of the diffusion-reaction path implies that diffusive flow and chemical flux are constant throughout this path, i.e. when we consider flow in direction $IV\to I\to II\to III\to IV$ we obtain
\begin{equation}
J_d=J_{I\rightleftharpoons II}=J_p=J_{III\rightleftharpoons IV} \equiv J\;.\label{flowconservation}
\end{equation}
   
We will now derive the steady state proton flow $J$, and hence steady state rotation. This requires determination of the steady state probability densities at the transition points $\rho_I=\rho_d(x_I),\;\rho_{II}=\rho_p(x_{II}),\;\rho_{III}=\rho_p(x_{III}),\;\rho_{IV}=\rho_d(x_{IV})$. On the diffusion paths the respective gradient of these probability densities $\rho_A-\rho_B$ with $(A,B)=(II,III),\; (IV,I)$, maintains the corresponding diffusive flow. By generalization of Fick's law for gradient driven diffusion we could recently derive this flow compactly as a function of occupation probability in, and first passage time through the path \cite{bauer2005}. In detail: for a perfect absorbing boundary $\rho_B=0$, unidirectional flow in the steady state is
\begin{equation}
J_{A\to B}=\frac{n_{A\to B}}{\tau_{A\to B}}\;\rho_A\;,\label{flowuni}
\end{equation}
and vice versa for $B\to A$. Here 
\begin{equation}
\tau_{A\to B}=D^{-1}\int_{x_A}^{x_B}dx \;e^{\Phi(x)}\int_{x_A}^{x} d\xi \;e^{-\Phi(\xi)}\label{fptuni}
\end{equation} 
is the regular mean first passage time \cite{szabo}, and 

\begin{equation}
n_{A\to B}=\rho_A^{-1}\int_{x_A}^{x_B} \rho(x)dx \label{occupation}
\end{equation}
is the specific occupation number, which by $\rho_A\;n_{A\to B}$ defines the probability to find the system within the diffusion path. This number is independent from the boundary value $\rho_A$, as long as non self interacting diffusing systems are considered, which is the case in our model (see Eq.~(\ref{Smoluchowski}). Bidirectional steady state flow for arbitrary values $\rho_A,\; \rho_B$, is simply the superposition of unidirectional flows, $J_{A\rightleftharpoons B}=J_{A\to B}+J_{B\to A}$. Since this flow vanishes for equal boundary densities \cite{commentconservative}, Eq.~(\ref{flowuni}) implies $n_{A\to B}/\tau_{A\to B}=n_{B\to A}/\tau_{B\to A}$. The generalized macroscopic Fick's diffusion law then reads
\begin{equation}
J_{A\rightleftharpoons B}=\frac{n}{\tau}\;\big(\rho_A-\rho_B\big)\;,\label{flow1}
\end{equation}
with symmetrized first passage time $\tau$  and specific occupation number
$n$ \cite{linear}
\begin{eqnarray}
\tau&=&\frac{1}{2}\;(\tau_{A\to B}+\tau_{B\to A})\label{symmtau}\\
n&=&\frac{1}{2}\;(n_{A\to B}+n_{B\to A})\label{symmn}\;.
\end{eqnarray}
Both parameters depend on the interaction $\Phi$ by \cite{bauer2005, BauerNadlerpnas}
\begin{eqnarray}
n&=& \frac{L}{2}\; \langle e^{-\Phi(x)}\rangle \label{n}\\ \tau&=&\frac{L^2}{2 D}\;\langle e^{-\Phi(x)}\rangle\;\langle
e^{\Phi(x)}\rangle \label{tau}
\end{eqnarray} 
where the brackets denote the spatial average $\langle\rangle=1/L\int_{0}^{L}$, with $L$ and $D$ are the  length and diffusion coefficient on the respective path. 

When we set diffusive flows of Eq.~(\ref{flow1}) and chemical fluxes of Eqs.~(\ref{flow3}) according to 
Eq.~(\ref{flowconservation}) equal $J$, we obtain a set of four linear equations for $\rho_I,\cdots,\rho_{IV}$.
\begin{equation}
\left.\begin{matrix} \left.\right.&\left.\right.&J_{I\rightleftharpoons II}&= & k_+\;c_1\;\rho_I - k_-\;\rho_{II} \\ J_p&=&J_{II\rightleftharpoons III} &=&n_p/\tau_p(\rho_{II}-\rho_{III})\\ \left.\right.&\left.\right.&J_{III\rightleftharpoons IV}&=& k_-\;\rho_{III}-k_+\;c_2\;\rho_{IV}\\J_d&=&J_{IV\rightleftharpoons I} &=&n_d/\tau_d(\rho_{IV}-\rho_{I})   \end{matrix}\right\rbrace=J\label{linearequations}
\end{equation}

 In addition conservation of probability (Eq.~(\ref{conservationprobability1})) to find the
system within some state must hold, i.e. when expressed in terms of specific occupation probabilities (Eq.~(\ref{occupation})) this reads
\begin{equation}
\rho_I\; n_{I\to IV}+\rho_{II}\; n_{II\to III}+\rho_{III}\; n_{III\to II}+
+\rho_{IV}\; n_{IV\to I}=1\;.\label{conservationprobability}
\end{equation}
So in summary we have five linear equations from which the four probability densities at the transition points $\rho_I,\cdots,\rho_{IV}$ and steady state flow $J$ can be obtained.
When we assume that the protonation / de-protonation reaction rate is much faster than diffusive motion of the motor, i.e. access of protons from the bulk to the rotor is not limiting rotation \cite{Feniouk}, flow is obtained as
\begin{widetext}
\begin{equation}
J=\frac{\frac{1}{2}\;K\;\left(c_1-c_2\right)}{\tau_p + K c_1\; K c_2\; \tau_d +
K(c_1+c_2)\;\frac{\tau_p+\tau_d}{2}+
K(c_1-c_2)\;\frac{(\Delta\tau_p+\Delta\tau_d)}{2}}\;,\label{flowmotor}
\end{equation}
\end{widetext}
with 
\begin{equation}
\Delta \tau=\frac{1}{2}\; (\tau_{A\to B}-\tau_{B\to A})\;,
\end{equation}
 as the asymmetric counterpart of the symmetric first passage time $\tau$ (Eq.~\ref{symmtau}). $\Delta\tau$ quantifies the asymmetry of the interaction $\Phi(x)$, and vanishes when $\Phi(x)$ is symmetric on the diffusion path. The generalized  equilibrium constant 
\begin{eqnarray}
K&=&\frac{k_+}{k_-}\; \frac{n_p}{n_d}\cr\cr
&=& e^{-g_0}\frac{\langle e^{-\Phi_p(x)}\rangle}{{\langle e^{-\Phi_d(x)}\rangle}}\;\frac{L_p}{L_d}\;,\label{defK}
\end{eqnarray}
comprises the ring-neighborhood interaction $\Phi$ and the standard free energy of protonation 
\begin{equation}
g_0=-\ln(k_+/k_-)\;.
\end{equation}
Note that the latter is related to the acid dissociation constant of the proton carrier by $g_0\approx -2.3\; pK_a$. The term equilibrium constant becomes evident, when one considers the system under equilibrium conditions, i.e. $c_1=c_2=c$, and hence $\rho_I=\rho_{IV}$, $\rho_{II}=\rho_{III}$, and $(k_+/k_-) c=\rho_{II}/\rho_I$. Then the definition of the specific occupation number implies that  $K c$ gives is the ratio of the occupation probabilities $N=\rho n$, i.e. $K c=N_p(c)/N_d(c)$. Thus,  
procedures which favor the probability to find the system in the protonated state, e.g. lowering $g_0$ or an increase/decrease of binding strength in the protonated/deprotonated ($\Phi_p\downarrow,\; \Phi_d\uparrow)$ state, increase $K$ and vice versa.   

\begin{figure}
\label{figure3}
\includegraphics[width=8cm, angle=-0]{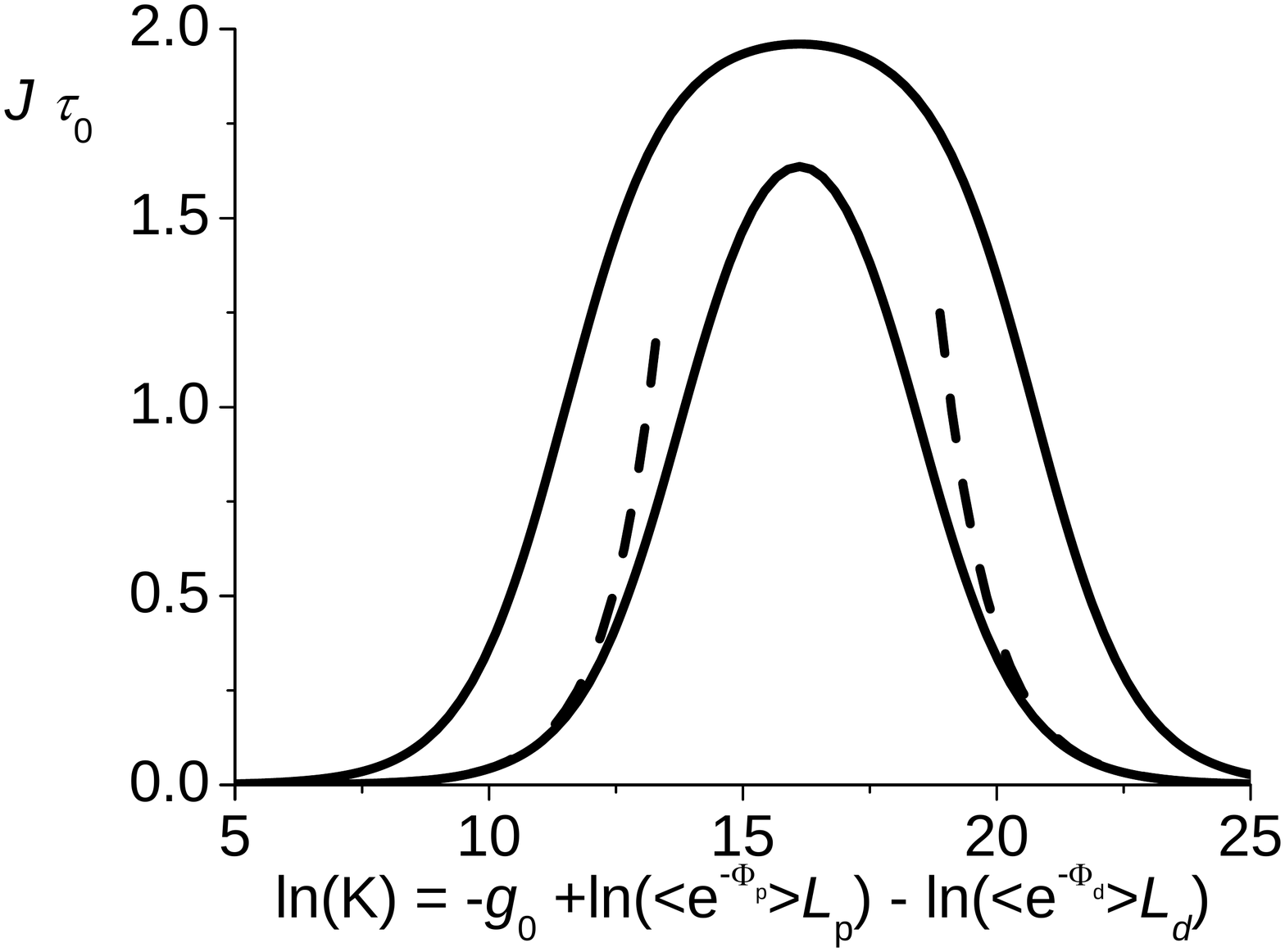}
\caption{Proton flow as a function of the constant $K$ (see Eq.~(\ref{defK})), which comprises ring-neighborhood interaction and proton binding strength of the proton shuttle (protomer). The protonated and deprotonated state are assumed to have identical diffusion properties ($L_p=L_d$, diffusion constant $D$), and potentials are constant, i.e. $\tau_p=L_p^2/2D=\tau_d=L_d^2/2$. Flow is normalized to the first passage time $\tau_0=(L_p+L_d)^2/(2D)$, i.e. the time needed to pass the whole diffusion length $L=L_p+L_d$. Two scenarios are shown, $c_1=10^{-6},\;c_2=10^{-8}$ (narrow curve) and $c_1=10^{-5},\;c_2=10^{-9}$ (wide curve).  Note that according to Eq.~(\ref{Kmax}) maximum flow is located at $K^{-1}=\sqrt{c_1 c_2}=10^{-7}$. The dashed curves are the approximations according to Eqs.~(\ref{highc}) for the first scenario.}
\end{figure}

In the limiting cases of low and high concentrations Eq.~(\ref{flowmotor}) has the form of a Fick's or inverse Fick's diffusion law respectively, i.e.
\begin{eqnarray}
J_{low}&\approx& \frac{1}{2}\;\tau_p^{-1} K\; (c_1-c_2)\;,\cr\cr J_{high}&\approx&\frac{1}{2}\;\tau_d^{-1}
K^{-1}\; (c_2^{-1}-c_1^{-1})\;.\label{highc}
\end{eqnarray}
Hence, increasing $K$, by increasing the probability to find the rotor in the protonated state increases flow in the setting of low proton concentrations, and vice versa for high proton concentrations. This implies an optimum ratio of occupation probabilities, reflected by $K$, for which flow, and hence rotation speed,  reaches a maximum (Fig.~3). For symmetric interactions, i.e. $\Delta\tau=0$, Eq.~(\ref{flowmotor}) determines this as 
\begin{equation}
K_{max}=\sqrt{\tau_p/\tau_d}\;\;1/\sqrt{(c_1\;c_2)}\;.\label{Kmax}
\end{equation}

The above considerations may be formulated in a very elegant way in terms of potentials, which allows generalization to the case of external workload. By defining the potentials of the respective driving forces as \begin{equation}
G_i=-\ln(K c_i)\;, \label{defG}
\end{equation}
we obtain 
\begin{widetext}
\begin{equation}
J=\frac{1}{2}\;\frac{\sinh(-\Delta G/2)}{\overline{\tau}_a\cosh(-\Delta G/2)+\overline{\Delta\tau}_a\sinh(-\Delta G/2)+\overline{\tau}_g \cosh\left((G_1 + G_2-\ln(\tau_d/\tau_p))/2\right)}\;,  \label{flowpotentials}
\end{equation}
\end{widetext}
with $\Delta G=G_1-G_2=-\ln(c_1/c_2)$ as the free energy gained during one cycle on the reaction diffusion path (Fig.~2), and $\overline{X}_a=(X_p+X_d)/2$ as arithmetic, and $\overline{X}_g=\sqrt{X_p X_d}$ as  geometric mean values of the variable $X$ over the protonated and deprotonated states. It should be mentioned that Eq.~(\ref{flowpotentials}) also holds when the standard free energy of protonation differs in the channels $g_0^{(1)},\; g_0^{(2)}$. This situation occurs when not only a chemical but an electro-chemical gradient is present with $\Delta g_0$ as the membrane potential. According to Eqs.~(\ref{defK}, \ref{defG}) this implies two constants $K^{(1)},\;K^{(2)}$ and corresponding potentials $G_i=-\ln(K^{(i)}\;c_i)$, with the free energy difference $\Delta G=-\ln(c_1/c_2)+\Delta g_0$.     

With the above Equation one can now determine the optimum relation of ring neighborhood interaction quantified by $\Phi$ and standard free energy of protonation $g_0$ to achieve maximum flow, i.e. rotation. In a first step, we  vary the free energy profile without changing the first passage times. This can be achieved by shifting  the standard free energy of protonation $g_0$, and/or according to Eqs.~(\ref{fptuni}-\ref{tau}) by constant shifts of $\Phi_p$ and $\Phi_d$. Equation~(\ref{flowpotentials}) then predicts flow maximum for  
\begin{equation}
G_1+G_2=\ln(\tau_d/\tau_p)\,. \label{condoptflow}
\end{equation}
The impact of this condition for the free energies $G_i$ becomes best evident when first passage times in the particular protonation states are identical, i.e. $G_1=-G_2$. Instead of taking $G_1$ and $G_2$  as opposing driving forces one can interpret $G_1$ and $-G_2$ as synergistic driving forces, i.e. maximum flow occurs when driving forces cooperate optimally. For non identical first passage times, one has to adjust by 
$\ln(\tau_p/\tau_d)$.    

We now vary the first passage times and simultaneously keep the condition (\ref{condoptflow}) fulfilled by appropriate shifts of free energy components. The Cauchy-Schwarz inequality then states for the symmetrized first passage time (Eq.~(\ref{tau})) that
$\tau_i\ge L_i^2/(2D)$ ($i=p,d)$, i.e. the first passage time reaches its minimum value when the interaction potentials become constant on the respective diffusion paths. Since then $\overline{\Delta\tau}$ vanishes, flow in Eq.~(\ref{flowpotentials}) reaches its maximum 
\begin{equation}
J_{max}=\frac{1}{2}\;\frac{\sinh(-\Delta G/2)}{\overline{\tau}_a\cosh(-\Delta G/2)+\overline{\tau}_g }\;,  \label{flowpotential}
\end{equation}

\section*{With External Workload}
\begin{figure}
\label{figure4}
\includegraphics[width=8cm, angle=-0]{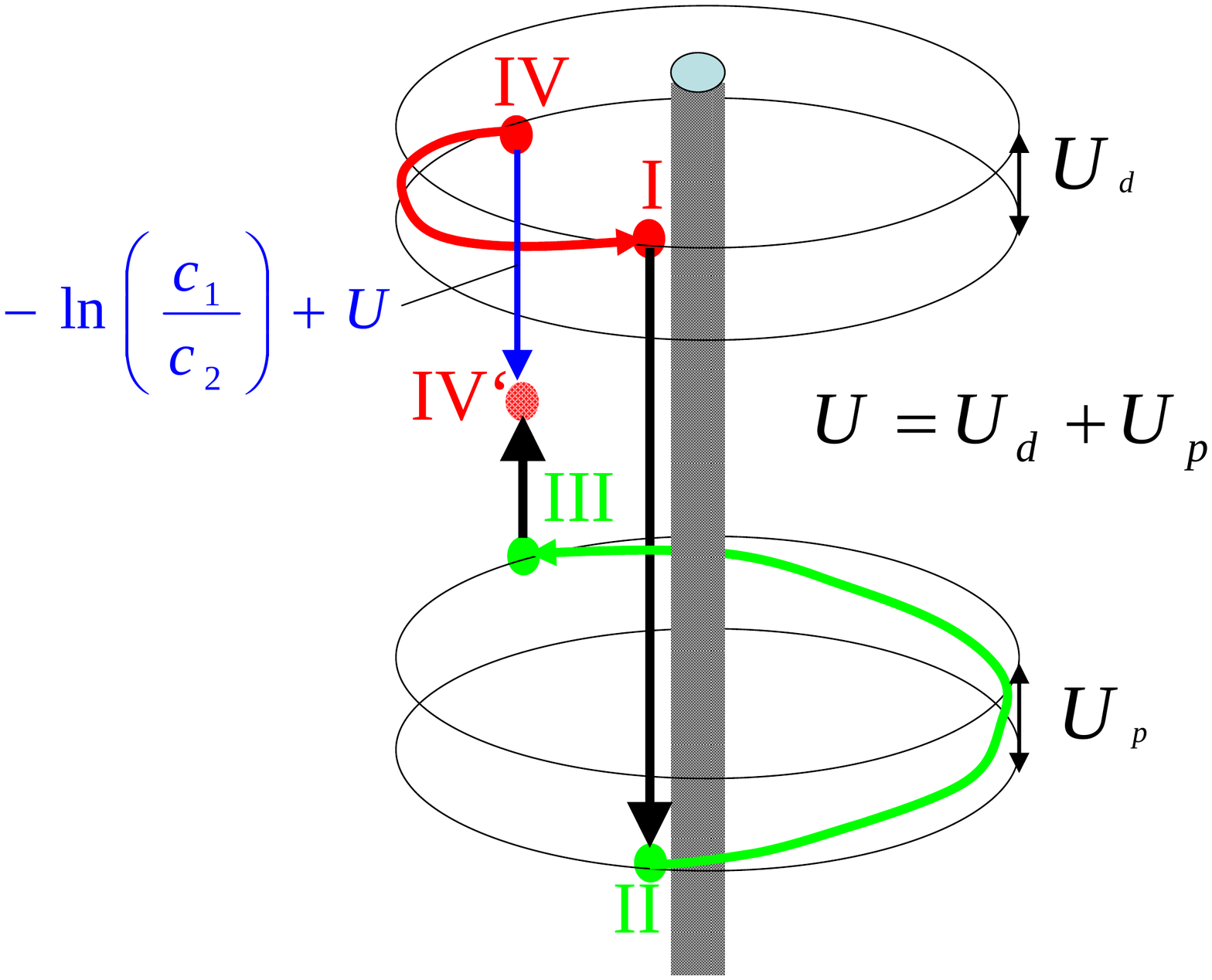}
\caption{Free energy landscape for a motor working against constant external workload. For simplicity the free energy related to internal ring neighborhood interactions $\Phi$ is not shown. We assume that high energy barriers, acting as check valves, confine the motor to the left diffusion path in the deprotonated and to the right in the protonated state. The motor has to afford the work $U_d=-FL_d$ in the deprotonated and $U_p=-F L_p$ in the protonated state, i.e. after one complete turn $U=U_p+U_d$, by gaining the free energy $\Delta G+U=-\ln(c_1/c_2)+U$.}
\end{figure}

The motor is now supposed to works against an external constant force $F$. As in the previous section, we assume that high energy barriers confine deprotonated and protonated state to complementary diffusion paths (see insert in Fig.~2). The free energy landscape is that of the previous section with an additional potential term $-F\;x$ which accounts for the external workload. After one cycle the external work is $U=U_p+U_d=- F (L_p+L_d)$ (Fig.~4).   
Flow is determined as in the previous section (Eq.~\ref{flowpotentials}) after adjusting parameters for the external workload (see Appendix). Free energies are adjusted to $G_1+G_2\to G_1+G_2+U$ and the free energy gained after one cycle $\Delta G \to \Delta G+U=-\ln(c_1/c_2)+U$,
\begin{equation}
J=\frac{1}{2}\frac{\sinh\left(-\frac{\Delta G + U}{2}\right)}{\overline{\tau}_a\cosh\left(\frac{\Delta G + U}{2}\right)+\sinh\left(-\frac{\Delta G + U}{2}\right)\overline{\Delta\tau}_a+\overline{\tau}_g\cosh\left(\frac{G_1+G_2+U-\ln(\tau_d/\tau_p)}{2}\right)}\label{flowgeneralU}
\end{equation}    
We focus on simple constant ring interactions $\Phi_i$ in the respective diffusion paths, which guarantees that the first passage times remain constant when interactions are varied (Eq.~\ref{fptuni}). The potential $G$ may then be decomposed into a component related to the external force and one related to the internal (ring neighborhood) interactions and chemical transitions, $G_i=G_{ext}+G_{i,int}$, with (see Appendix)
\begin{eqnarray}
 G_{i,int}&=&\Phi_p-\Phi_d+g_0-\ln(c_i)\;,\label{Gint} \\
 G_{ext}&=&-\ln[(1-e^{-U_p})/(1-e^{-U_d})]\label{Gext}\;.
 \end{eqnarray}
This separation makes the interdependence of ring neighborhood interaction $\Phi$ and standard free energy $g_0$ concerning their effect on the driving forces $G_{i,int}$ evident. Simulations of the F0 ring motor working against at constant workload by Elston \cite{Elston} demonstrated for a stronger binding in the deprotonated state, i.e. a decrease of $\Phi_d$, that maximum rotation rate occurred at a stronger proton binding value, i.e. a decrease of $g_0$. This is directly predicted  by the form of $G_{int}$, since a balanced shift of interaction and standard free energy keeps $G_{int}$ constant, i.e. in this case the driving force at which maximum rotation rate occurred.     

Figure 5 demonstrates that increasing workload $U$ reduces flow  and shifts the maximum toward interactions favoring the protonated state. This is evident, since maximum flow occurs for 
\begin{equation}
G_1+G_2+U=\ln(\tau_d/\tau_p)\;,
\end{equation}
 i.e. at  
 \begin{equation}
 \Phi_p-\Phi_d+g_0=-G_{ext}+\frac{1}{2}[-U+\ln(c_1c_2)+\ln(\tau_d/\tau_p)]\; \label{Phigmax}.
 \end{equation} 
\begin{figure}
\label{JvsPhi}
\includegraphics[width=9cm, angle=-0]{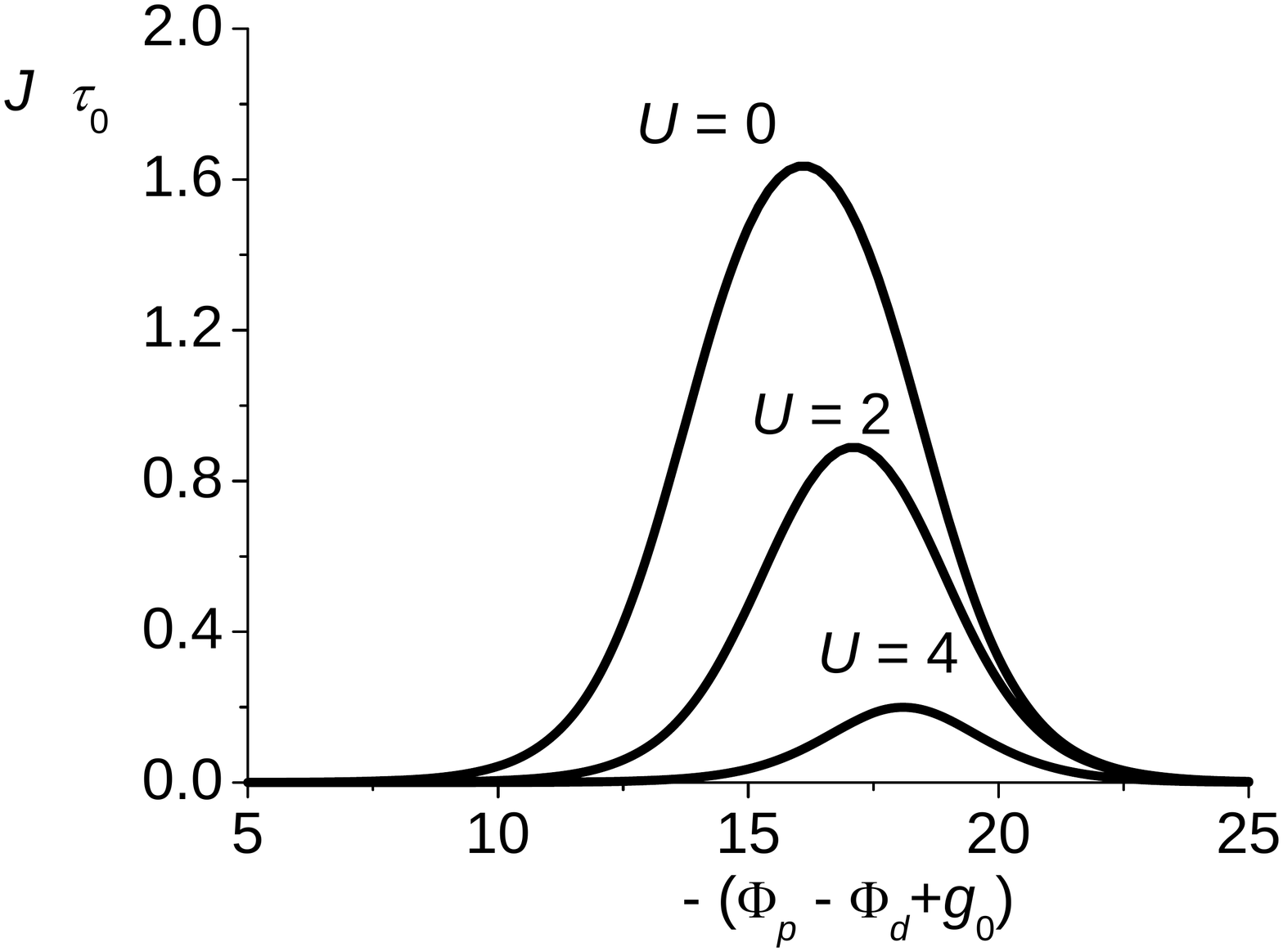}
\caption{Flow under different workload conditions $U$ as a function of ring-neighborhood interaction $\Phi$, and standard free energy of protonation $g_0$. Proton concentrations are $c_1=10^{-6},\;c_2=10^{-8}$. The geometric and diffusion parameters are that of Fig.~3, i.e. equal for the protonated and deprotonated state. This implies $G_{ext}=0$ (Eq.~(\ref{Gext})), i.e.  maximum flow occurs at $\Phi_p-\Phi_d+g_0=1/2(-U+\ln(c_1c_2))$, i.e. at  -16.1 , -17.1, and -18.1, (Eq.~(\ref{Phigmax})). }
\end{figure}

\begin{figure}
\label{figure6}
\includegraphics[width=9cm, angle=-0]{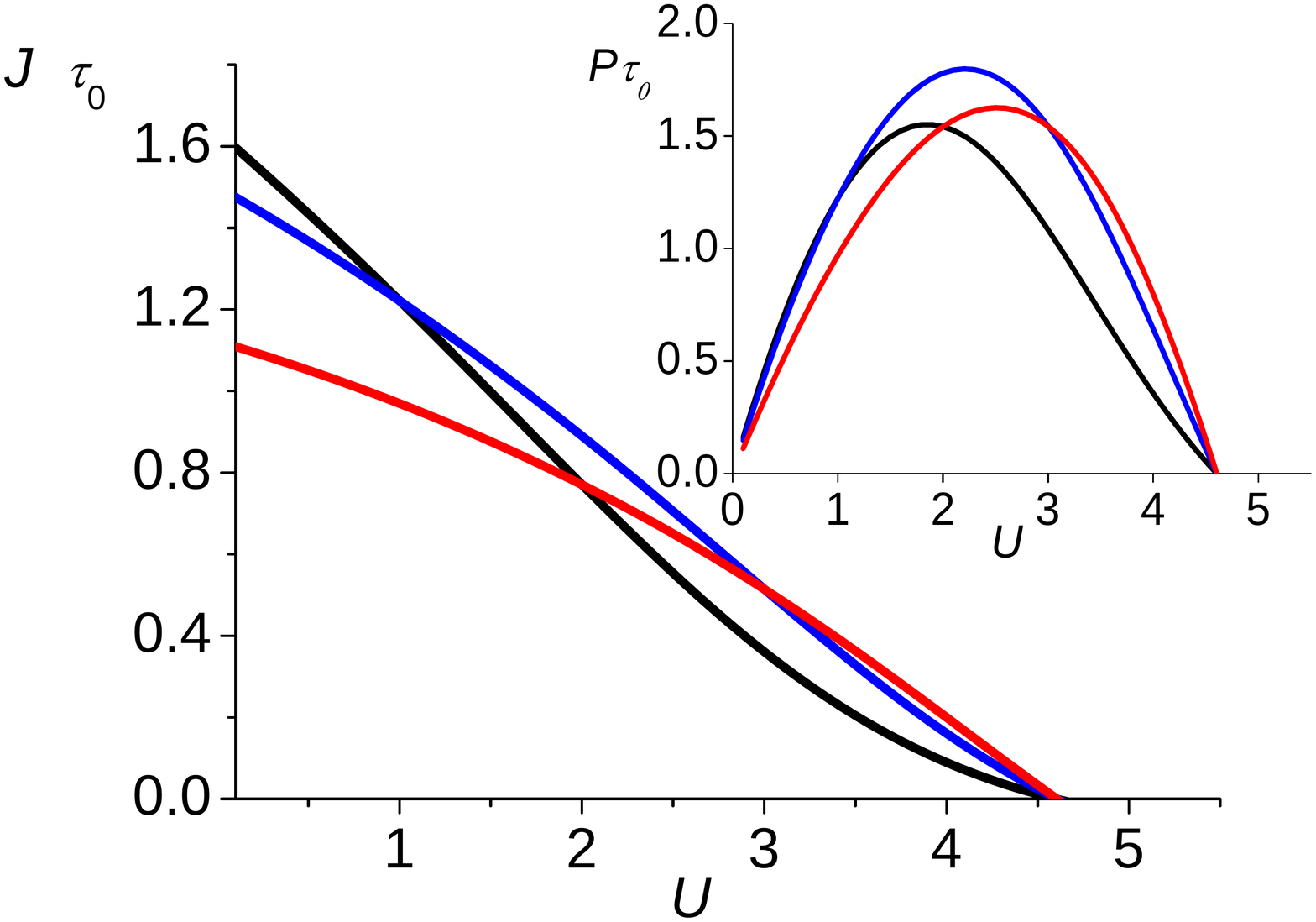}
\caption{Flow as a function of workload $U$ for different ring-neighborhood interactions 
$\Phi_p-\Phi_d+g_0=-16.1\; \mbox{(black)}, -17.1\; \mbox{(blue)}, -18.1\; \mbox{(red)}$ corresponding to maximum flow for $U=0,2,4$ (see previous Figure). At workload $U=4.6=\ln(c_1/c_2)=-\Delta G$, flow and hence motor action vanishes. The insert shows the corresponding performance (Eq.~(\ref{performance}))with the same color labeling.}
\end{figure}

When flow is considered as a function of workload (Fig.~6), one obtains a more and more parabolic shape of the curves, when  internal interactions favor the protonated state, i.e. $(\Phi_p-\Phi_d+g_0)$ decreases. This parabolic behavior implies that the motor even under higher workload conditions maintains sufficient performance.   
For determination of this motor performance the following consideration is useful.
When a single motor is at some position $x$, it performs the external mechanical work $dW=-F dx$ after moving to $x+dx$. For an ensemble of motors, which is in the steady state, the number of motors moving $dx$ at position $x$ is $J dt$, i.e. the local work  performed is $J dt (-F) dx$. Performance as work per time over the whole length is then 
\begin{eqnarray}
P&=& \int_0^{L}dx\; (-F)\;J\cr\cr
&=& U\;J\;.\label{performance}
\end{eqnarray}
Interestingly the motor may run at high performance even when workload increases as long as "tuning" by the interaction $(\Phi_p-\Phi_d+g_0)$ is appropriate (insert in Fig.~6).

 \section*{Motor without Check Valve Mechanism}
 \subsection*{No Workload}
 \begin{figure}
\label{figure7}
\includegraphics[width=9cm, angle=-0]{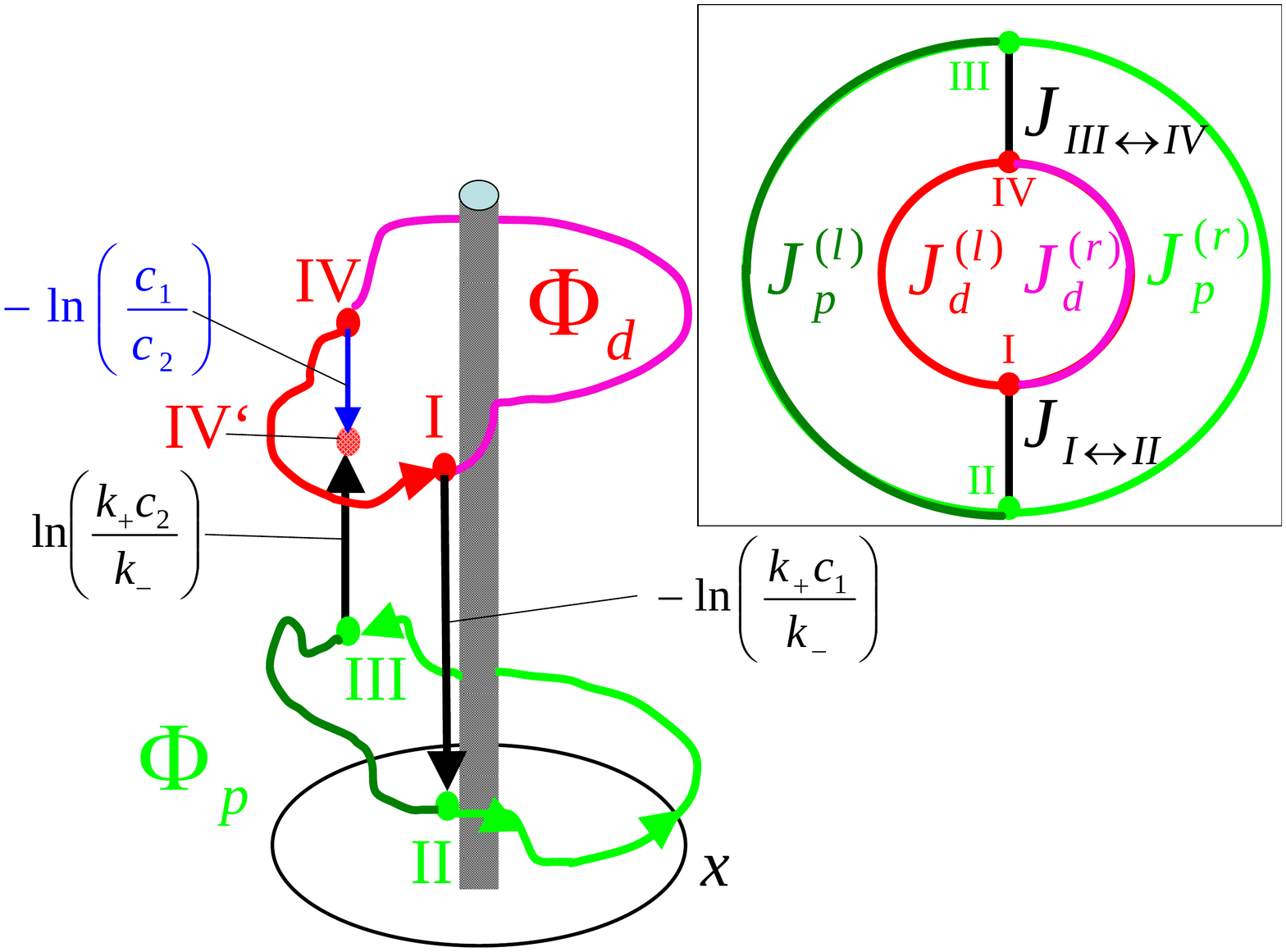}
\caption{Free energy profile of the rotor without workload but no check valve mechanism. The labeling is that of Fig.~2. Here, in neither protonation state the rotor is confined to sections of the diffusion path by infinite high energy barriers acting as check valves. Instead finite energy barriers (magenta in the deprotonated, dark green in the protonated state) only favor sections of the diffusion path, the left (red) in the deprotonated, the right (bright green) in the protonated state. This favors the rotor to run in counter clockwise direction and this preferred way is marked by arrows. The insert shows the topology of the diffusion-reaction pathways and the corresponding diffusive flows and chemical fluxes.}
\end{figure}

We now consider the rotor without the constraint that infinite high energy barriers confine diffusive motion to complementary sections of the diffusion path in respective protonation states (Fig.~7). This implies that the rotor has the option to move in both directions at the cross points of chemical transitions and diffusion paths.  When for example the rotor has undergone the chemical transition (protonation) $I\to II$, it may proceed its way in the protonated state either in counterclockwise (right) or clockwise (left) direction. The gain of free energy ($-\ln(c_1/c_2)$) after one completed cycle is the same for both ways. Hence, it is important to stress, that  backward rotation, which in our example in Fig.~7 is clockwise, is not just a phenomena of fluctuations, but a ''thermodynamically allowed'' way to gain free energy. From the view of the protons this backward rotation can be interpreted as leak flow, which reduces the coupling between overall flow proton flow and directed rotation.  

In the steady state the conservation of flow holds when chemical flux is distributed into diffusive flows on the left and right path and vice versa (insert in Fig. 7). In cyclic direction (see also Eq.~(\ref{flowconservation})) this reads  
\begin{equation}
J_{I\rightleftharpoons II}=\underbrace{J_p^{(l)}+J_p^{(r)}}_{=J_p}=J_{III\rightleftharpoons IV}=\underbrace{J_d^{(l)}+J_d^{(r)}}_{=J_d} \equiv J\;.\label{flowconservation2}
\end{equation}
Since the left and right diffusive flows refer to rotation in opposite direction, the effective rotation is 
\begin{eqnarray}
J_{rot}&=&J_d^{(l)}-J_p^{(l)}\cr
&=&J_{p}^{(r)}-J_d^{(r)}\label{rotflow}
\end{eqnarray}
where the latter relation follows from Eq.~(\ref{flowconservation2}). 
One can define the fraction $f$ of overall flow $J$ which is directed into effective rotation
\begin{equation}
f=J_{rot}/J=(J_d^{(l)}-J_p^{(l)})/(J_p^{(l)}+J_p^{(r)})\;,\label{efficiency}
\end{equation}
which can also interpreted as the coupling strength of rotation to proton flow. In the absence of workload the relation $|f|\le 1$ holds, since the conservative drift forces  $F_i(x)=-\Phi_i'(x)$ conserve the direction of diffusive flows ($J_i^{(l)},\;J_i^{(r)}$) within a protonation state $i=d,\;p$. This direction is then solely determined by the probability gradient i.e. in our setup with $c_1>c_2$ it points from $II\to III\; \text{and}\; IV\to I$. Hence,  
$|J_d^{(l)}-J_p^{(l)}|\le \max(|J_d^{(l)}|,\;|J_p^{(l)}|)\le |J|$. The  relation $|f|\le 1$ does not hold generally for  non-conservative forces, e.g. when a constant external workload is present. Depending on its strength this force may rotate the motor within the protonated state, and by this decouple rotation $J_{rot}$ from proton flow $J$, a situation we will discuss below. 

Since in the steady state the generalized Fick's diffusion law (\ref{flow1}) holds for the particular diffusive flow components on the left or right paths, 
\begin{equation}
J^{(l,r)}_{A\rightleftharpoons B}=n_i^{(l,r)}/\tau_i^{(l,r)}(\rho_A-\rho_B)=J_i^{(l,r)}\;\label{flowpath}
\end{equation}
with $(A,B)=(IV,I),\;(II,III)$ and $i=p,\;d$, one obtains with Eqs.~(\ref{n}, \ref{tau}),
\begin{eqnarray}
f&=&\left(1+\frac{n_d^{(r)}}{n_d^{(l)}}\frac{\tau_d^{(l)}}{\tau_d^{(r)}}\right)^{-1}-\left(1+\frac{n_p^{(r)}}{n_p^{(l)}}\frac{\tau_p^{(l)}}{\tau_p^{(r)}}\right)^{-1}\cr\cr
&=&\left(1+\frac{L_d^{(l)}}{L_d^{(r)}}\frac{\langle e^{\Phi_d^{(l)}}\rangle}{\langle e^{\Phi_d^{(r)}}\rangle}\right)^{-1}-\left(1+\frac{L_p^{(l)}}{L_p^{(r)}}\frac{\langle e^{\Phi_p^{(l)}}\rangle}{\langle e^{\Phi_p^{(r)}}\rangle}\right)^{-1}\;,\label{re}
\end{eqnarray}  
where $L^{(l)},\;L^{(r)}$ are the lengths of the left and right diffusion paths, and $\langle\rangle$ denotes  the spatial average on the particular paths.
There is no rotation $f=0$, when free energy landscapes of the protonation states are congruent, i.e.
$L_d^{(l)}\langle e^{\Phi_d^{(l)}}\rangle/(L_d^{(r)}\langle e^{\Phi_d^{(r)}}\rangle)=L_p^{(l)}\langle e^{\Phi_p^{(l)}}\rangle/(L_p^{(r)}\langle e^{\Phi_p^{(r)}}\rangle$. Conversely free energy profiles favoring complementary paths in respective states synergistically increase rotatory efficiency $|f|\to 1$, e.g. as in Fig.~(7) a barrier on the right of the deprotonated state $\Phi_d^{(r)}\uparrow$, and left of the protonated state $\Phi_p^{(l)}\uparrow$. Infinite high barriers imply maximum efficiency $f=1$ since they act as check valves, a situation discussed in the previous section. In the real motor (Fig.~1) the synergism of free energy profiles is realized by the high hydrophilic-hydrophobic energy barrier which confines deprotonated protomers to the hydrophilic stator region. Since there is at least one deprotonated protomer in either protonation state, backward rotation is impeded, i.e. $f=1$. However, the situation would be different for motors in which the maximum number of deprotonated protomers facing the hydrophilic stator is one. In this case there would be no mechanism preventing the rotor from backward rotation in the protonated state.

The steady state flow  $J$ can be derived as in the previous section, based on conservation of flow (Eq.~(\ref{flowconservation}) and overall probability (Eq.~(\ref{conservationprobability}). To apply the Eqs.~(\ref{linearequations}) for this derivation, one has to formulate diffusive flows in the form of a generalized Fick's diffusion law Eq.~(\ref{flow1}). This is not restricted to diffusion on a linear path. Instead it holds for steady state diffusion through arbitrary domains, as long as the forces acting inside are conservative, i.e. when they derive from a potential $F(x)=-\Phi'(x)$ \cite{bauer2005}. Here the domain is no longer a simple path as in the previous sections, instead it consists of a left and right path in either protonation state as the insert in Fig.~7 shows: red/magenta in the deprotonated, green/dark green in the protonated state. The boundaries of the domains are the cross points $(A,B)=(IV,I),\; (II,III)$. The application of Eq.~(\ref{flow1}) requires the knowledge of the specific occupation number and first passage time. The specific occupation number is simply the sum of the specific occupation number of the left and right diffusion path. In both protonation state follows from Eq.~(\ref{n})
\begin{eqnarray}
n&=&n^{(l)}+n^{(r)}\cr
&=&L^{(l)}/2\; \;\langle \exp(-\Phi^{(l)})\rangle +L^{(r)}/2\;\; \langle \exp(-\Phi^{(r)})\rangle\cr
&=&(L^{(l)}+L^{(r)})/2\;\; \langle \exp(-\Phi)\rangle\label{nges}   
\end{eqnarray}
To obtain the first passage times one has to keep in mind that Eq.~(\ref{flow1}) holds for the particular flow components on either left or right diffusion path (Eq.~(\ref{flowpath})). Since $J_{d,p}=J_{A\rightleftharpoons B}=J^{(l)}_{A\rightleftharpoons B}+J^{(r)}_{A\rightleftharpoons B}$, and $n=n^{(l)}+n^{(r)}$, the inverse first passage time derives as the occupation probability weighted sum of the inverse first passage times on the left anf right diffusion path    
\begin{equation}
\frac{1}{\tau}=\frac{n^{(l)}}{n^{(l)}+n^{(r)}}\; \frac{1}{\tau^{(l)}}+\frac{n^{(r)}}{n^{(l)}+n^{(r)}}\; \frac{1}{\tau^{(r)}}
\label{taugesamt}
\end{equation}
Insertion of specific occupation number and first passage time from the above Equations in Eqs.~(\ref{flow1}, \ref{linearequations}), then provides as in the previous section flow $J$ from Eqs.~(\ref{flowmotor}, \ref{flowpotentials}).

\subsection*{With Workload}    
\begin{figure}
\label{figure8}
\includegraphics[width=9cm, angle=-0]{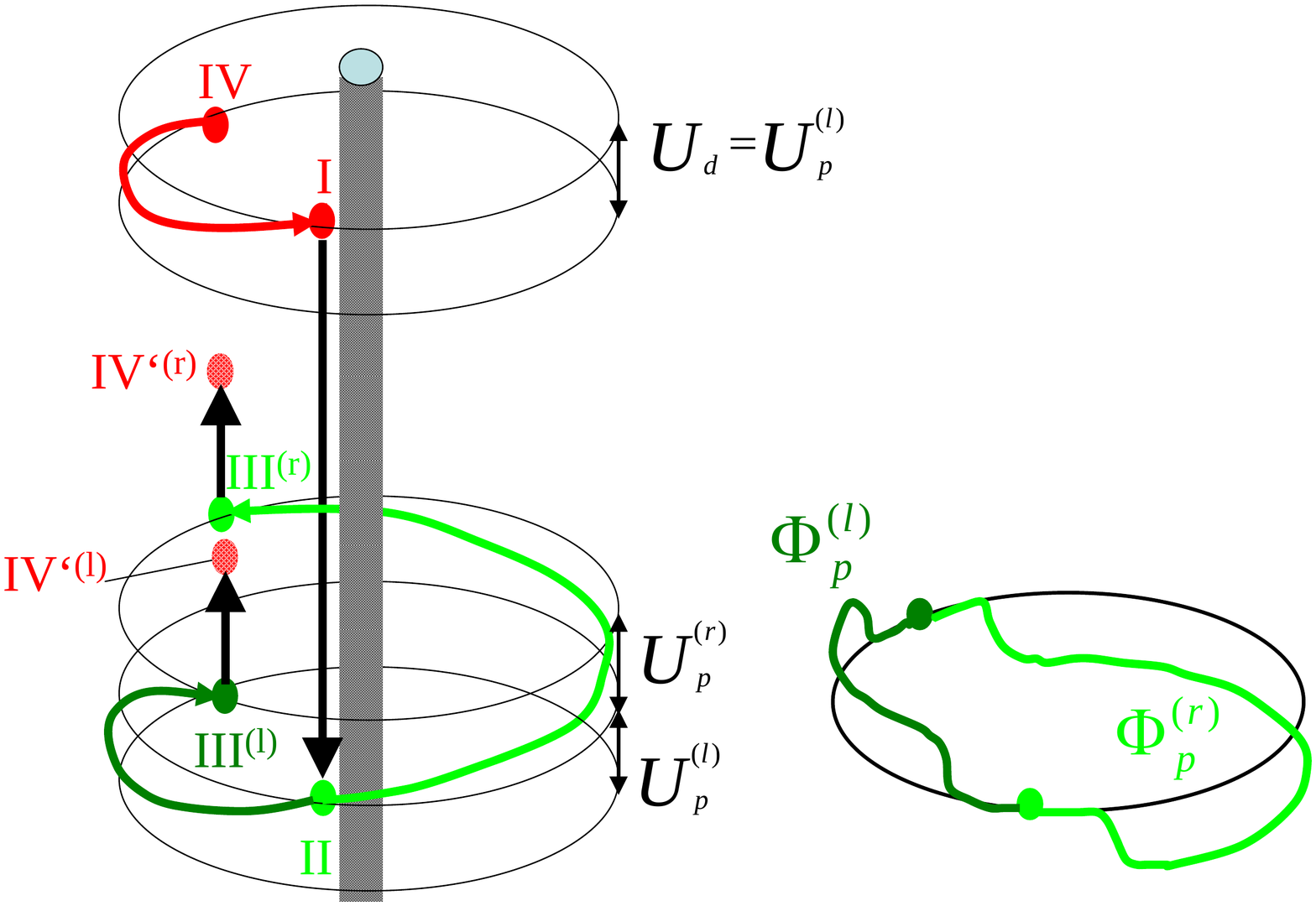}
\caption{Left: free energy topology for a motor working against constant external workload. A high energy barrier confines the motor to the left diffusion path in the deprotonated state. In this state, the motor first has to afford the work $U_d$ ($IV\to I$) before free energy is gained by protonation ($I\to II$). However in contrast to Fig.~5 the  motors has in the protonated state the choice either to complete the cycle in counterclockwise direction and to afford the work $U_p^{(r)}$, or it may rotate backwards and gain back the energy $U_d=U_p^{(l)}$. So the bifurcation of the reaction-diffusion path at point $II$ implies the option of two paths, by which the motor may reach its initial position ($IV'$) and gain free energy. For simplicity the internal interactions $\Phi_p$ are not shown on the left, but on the insert right which has to be superimposed on the free energy profile on the left.}  
\end{figure}

\begin{figure}
\label{figure9}
\includegraphics[width=8.5cm, angle=-0]{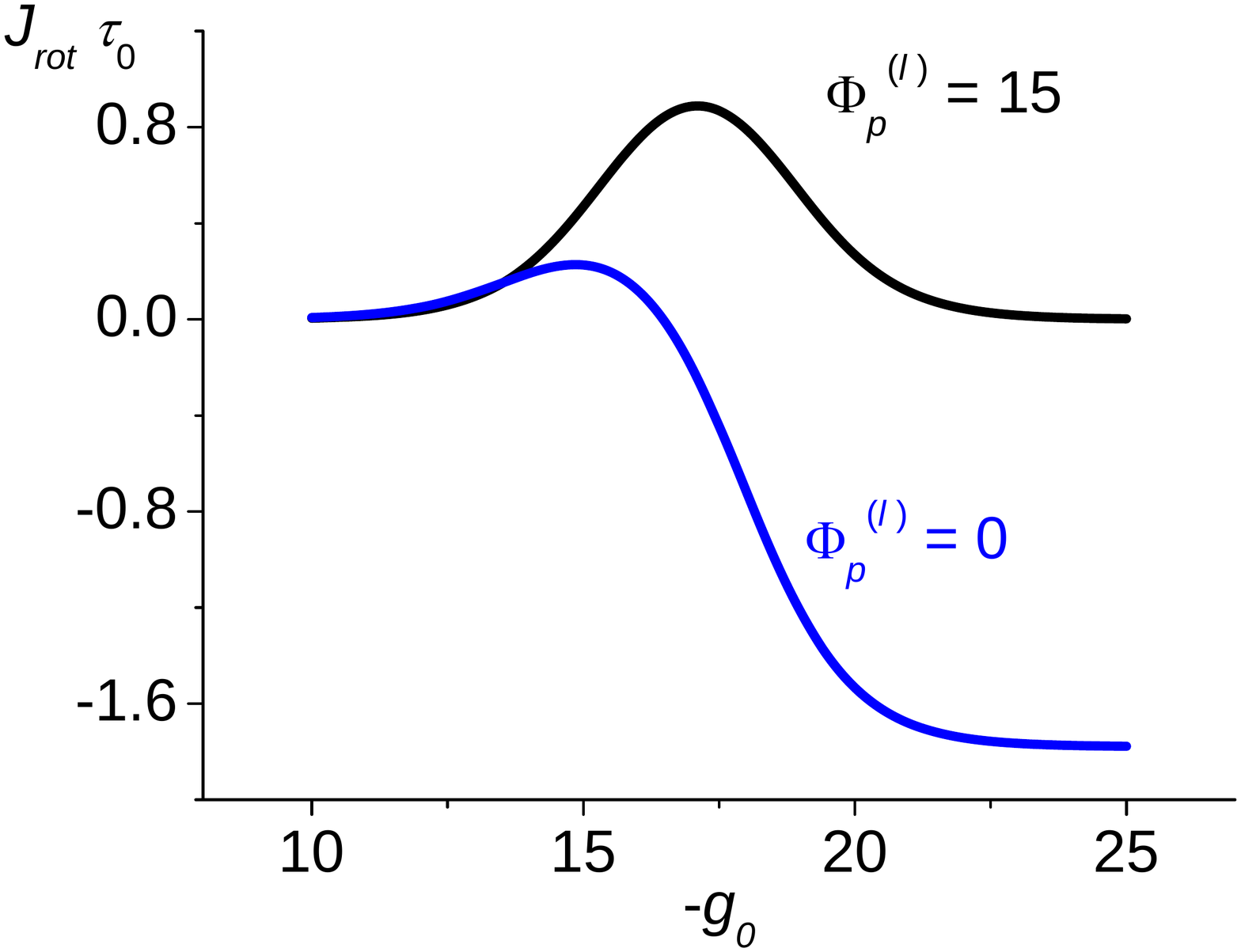}
\caption{Flow, which is effectively transformed into counterclockwise rotation,  i.e. against the external force, as a function of the standard free energy of protonation $g_0=-\ln(k_+/k_-)$. Concentration are $c_1=10^{-6}$ and $c_2=10^{-8}$, and potentials on the paths forming the   way in counterclockwise direction are $\Phi_d=\Phi_p^{(r)}=0$. The workload is $U=2$. A high energy barrier $\Phi_l=15$ prevents the motor from backward rotation (black line), whereas no barrier $\Phi_l=0$ (blue line) implies clockwise rotation $J_{rot}<0$ when the energy level of the protonated state is lowered ($g_0$ decreases). }
\end{figure}

\begin{figure}
\label{figure10}
\includegraphics[width=8.5cm, angle=-0]{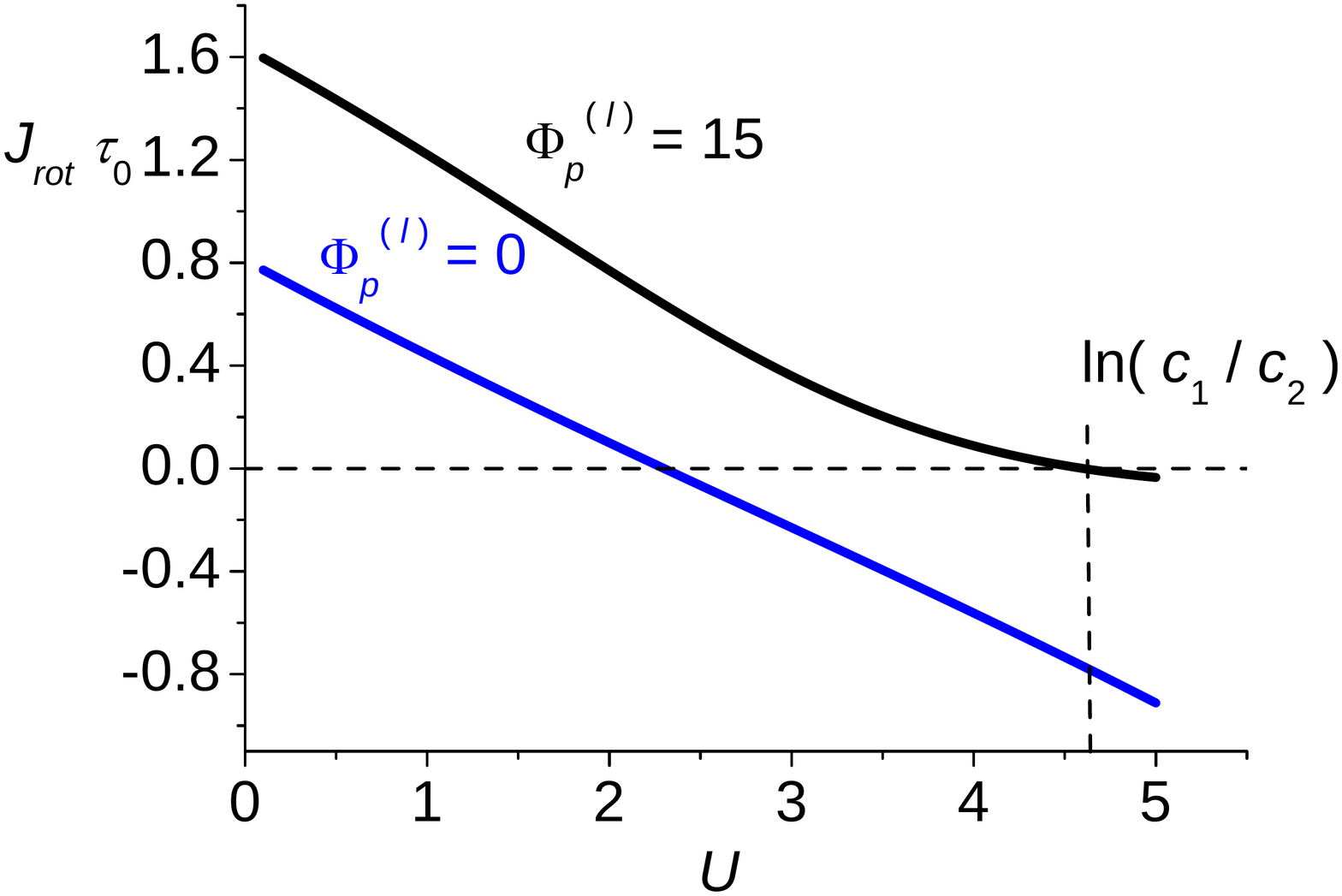}
\caption{Flow, which is effectively transformed into counterclockwise rotation $J_{rot}$ as a function of workload. Concentrations and potentials are as in the previous Figure. The standardized free energy of the protonation was chosen to be $g_0=-16.1$. With an energy barrier favoring counterclockwise direction ($\Phi_p^{(l)}=15$), the whole concentration gradient related free energy $|\Delta G|=\ln(c_1/c_2)=4.6$ may be converted into work. When there is no energy barrier ($\Phi_p^{(l)}=0$) the efficiency is much lower, as the zero point of this function (Eq.~(\ref{th})) demonstrates.}
\end{figure}

\begin{figure}
\label{figure11}
\includegraphics[width=8.5cm, angle=-0]{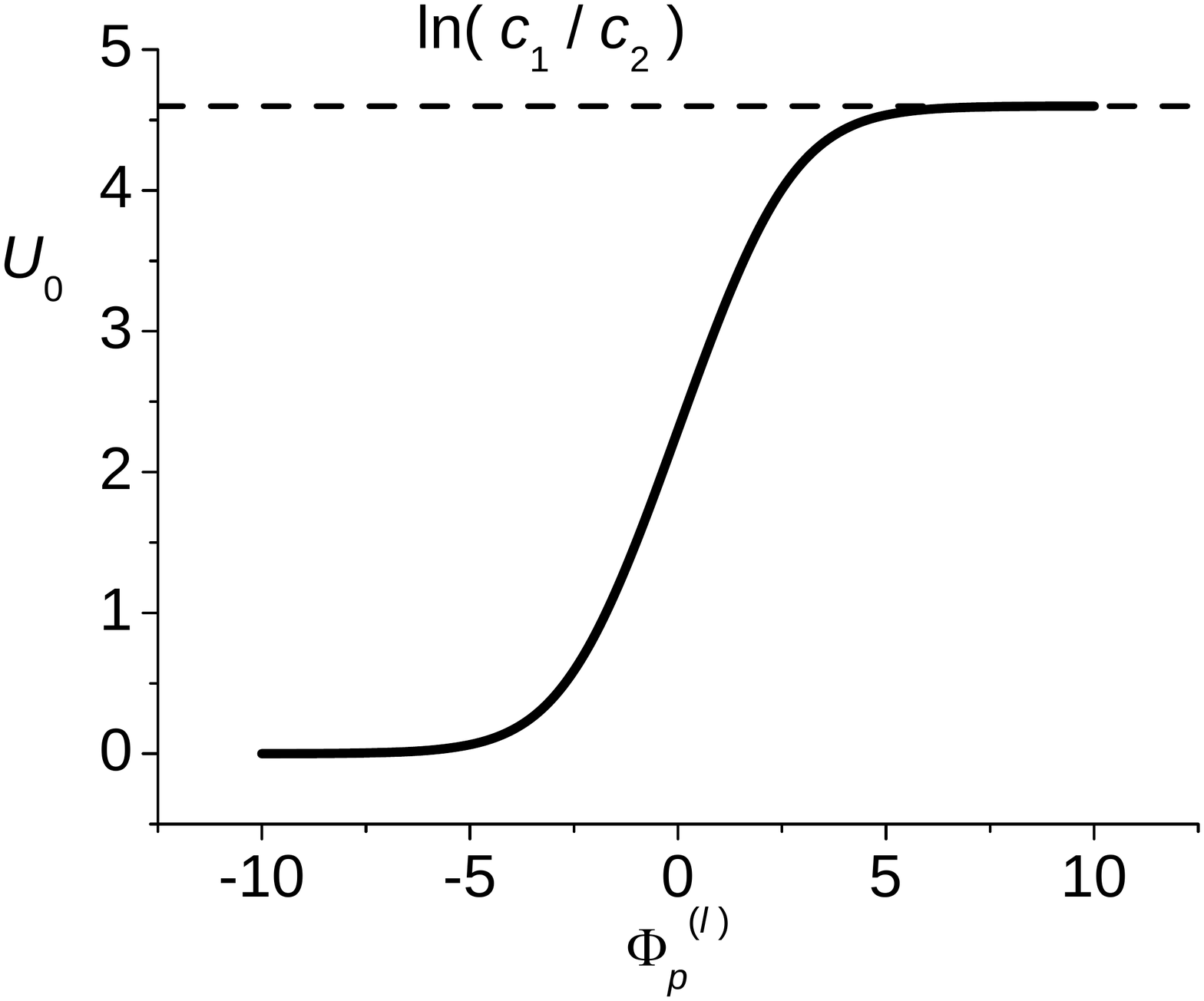}
\caption{Workload $U_0$ for which rotation vanishes as a function of the barrier $\Phi_p^{(l)}$ according to Eq.~(\ref{th}). Standard free energy is $g_0=-16.1$ and concentrations are $c_1=10^{-6},\; c_2=10^{-8}$. High barriers, which favor counterclockwise rotation against the external force ($\Phi_p^{(l)}\to +\infty$), make the workload approach the free energy difference $U_0\to |\Delta G|=\ln(c_1/c_2)=4.6$.}
\end{figure}

Now the rotor is now supposed to work against a constant force. For didactical reasons and to avoid too much complexity without gain of physical information we will not consider the topology of the reaction -diffusion path in its most general form as in Fig.~7. Instead we will confine the rotor by high energy barriers to the left path in the deprotonated state. In the protonated state it has the choice to take both ways (Fig.~8). In the presence of an external force this option has strong implications on the topology of the free energy landscape. When check valves were present this landscape would be just a simple, non-single valued function of a 1-D cyclic reaction diffusion path, i.e. after one turn the motor gains the free energy $\Delta G+ U$ (Fig.~4). However, now the motor has the option to diffuse either along the left or right path in the protonated state, i.e. it may gain the free energy $\Delta G+ U$ on the right vs. $\Delta G$ on the left path (Fig.~8). In other words this bifurcation of the reaction diffusion path leaves the option for the motor to proceed its way in clock- or in counterclockwise direction, however external work is only performed in the latter case. In addition gain of free energy is even higher when taking the "inefficient" path $\Delta G<\Delta G+U$. The effective rotational component as determined from Eq.~(\ref{rotflow}) is
\begin{eqnarray}
J_{rot}&=&J_d^{(l)}-J_p^{(l)}\cr
&=&J-J_p^{(l)}=J_p^{(r)};, \label{rotflowworkload}
\end{eqnarray}  
where we exploited that the motor is confined to the left diffusion path in the deprotonated state, i.e. $J=J_d^{(l)}$. The efficiency factor $f=(1-J_p^{(l)}/J)$ has not the simple form as in Eq.~(\ref{re}), since the external workload produces a non-conservative force field in the protonated state. For the determination of flow $J$ and its effective rotational component $J_{rot}$ see Appendix. 

We first consider the case that there is no energy barrier in the protonated state which could bias the direction of rotation, i.e. $\Phi_p^{(i)}=0$ (Fig.~9). Decreasing the standard free energy $g_0$ makes effective rotation first pass through a maximum and then vanish. At this point proton and leak flow are equivalent $J=J_p^{(l)}$ and flow on the right path in the protonated state is absent $J_p^{(r)}=0$ (Eq.~(\ref{rotflowworkload}). Further lowering of $g_0$ decreases the transition probability from the protonated to the deprotonated state and the rotor prefers to remain in the protonated state.
%, i.e. both states are almost decoupled. 
The rotor runs along the direction of the external force, $J_{rot}<0$, and by moving along the circular free energy landscape of the protonated state it gains the free energy $-U$ per round. We assume now constant interaction potentials on the left  $\Phi_p^{(l)}$  and right $\Phi_p^{(r)}$ section of the diffusion path in the protonated state, and in the deprotonated state $\Phi_d$. At the threshold value $g_0$ for which $J_{rot}$ vanishes, workload $U$ and ring neighborhood interactions satisfy (see Appendix)
\begin{equation}
e^{U}=\frac{1+e^{\Phi_p^{(l)}-\Phi_d+g_0-\ln c_2}}{1+e^{\Phi_p^{(l)}-\Phi_d+g_0-\ln c_1}} \;.\label{th}
\end{equation}
This Equation also makes the interdependence between standard free energy $g_0$ and the barrier $\Phi_p^{(l)}$ evident. The higher the barrier, the lower is the threshold for $g_0$ to make rotation vanish, i.e. a high barrier counteracts the external force. Interestingly the interaction on the right site in the protonated state $\Phi_p^{(r)}$ does not appear in the equation, i.e. it has no effect on the direction of rotation. One could argue that inversely to the barrier effect of $\Phi_p^{(l)}$, a high $\Phi_p^{(r)}$ should facilitate negative (clockwise) rotation in direction of the external force. For analysis let us first consider a situation in which the rotor runs against (counterclockwise) the external force, i.e. $J_{rot}>0$. Equation~(\ref{rotflowworkload}) then implies $J_p^{(r)}>0$. When we introduce a barrier on the right site by elevating $\Phi_p^{(r)}$ we reduce $J_p^{(r)}$ but it does not vanish, or become negative, since this barrier also works against the clockwise rotation in direction of the external force. In the limiting case $\Phi_p^{(r)}\to +\infty$, $J_p^{(r)}$, and hence $J_{rot}$ approach zero (Eq.~(\ref{rotflowworkload})), but they never become negative, i.e. there is no change of direction of rotation. So in summary $\Phi_p^{(r)}$ has no effect on the direction of rotation.           
The above Equation also defines the maximum workload $U_0$ for which rotation against the external force is maintained. A workload above this threshold implies backward rotation. When the barrier height   $\Phi_p^{(l)}$ is raised, this threshold increases, and for $\Phi_p^{(l)}\to +\infty$ it asymptocially reaches the free energy difference $U_0\to -\Delta G=\ln(c_1/c_2)$ (Fig.~10).

\section*{Discussion and Conclusion}

Whereas past interest of theorists mainly focused on the working principles of linear motors \cite{Fisher, Haenggi2002, Haenggi2005}, in this paper we presented a comprehensive analytical model of F0-like Brownian rotors, which are archetypes of rotatory biological molecular motors. These rotors transform a free energy difference maintained by an (electro)-chemical gradient into mechanical work by means of random Brownian motion. In contrast to the classical flashing ratchet concept \cite{Astumian}, in which on a molecular level chemical energy of the motor is directly transduced via a power strike mechanism into mechanical energy \cite{Wang1998, Yasuda}, these types of motors convert the free energy difference into entropic forces \cite{Parmeggiani, Astumian2, Wang}. This is realized by (chemical) transitions between different free energy profiles, the latter biasing motion into certain directions. Directed average motion results from different transition rates at different locations. In the case of the F0 rotor this is realized by different cation (proton) concentrations in respective access channels. Note that these transitions - though they may alter the free energy of the motor state - do {\it not} alter the mechanical energy at the point of transition \cite{Wang}. In contrast, the latter would be the basic mechanism of a classical flashing ratchet. 

To make a Brownian rotor run optimally requires both: The free energy difference must be converted into maximum  average motion, and - in addition - this average motion must be rectified into one direction, as in linear motors \cite{Suzuki}. We demonstrated that in our system free energy may be gained by forward as well as backward rotation. This is important since it demonstrates that backward movement is not only due to fluctuations. In our case it exists also as average motion on certain segments of the diffusion path - ie. the leak part - and is a consequence of the topology of the diffusion-reaction path and the overlying free energy profile. 

Theses considerations stress the importance of the interaction of the rotor with its surrounding. For F0-like rotors we derived analytical expressions for the dependence of flow on interactions, determining maximum flow for a given free energy difference. It could be demonstrated that even under high workload conditions optimum interactions implied a high performance of the motor.  
Furthermore the arrangement of interactions with respect to its rectifying property was analyzed. The latter determines the maximum workload the rotor can accomodate. However, only under perfect `` check valve '' conditions the maximum workload approaches the driving free energy difference. In its biological realization this situation is almost accomplished, since in either protonation state at least one deprotonated protomer is confined to the hydrophilic stator region (see Fig.~ 1). This problem of rectifying is typical for molecular rotors that work in the strong friction limit, i.e. when motion is over-damped and inertia effects are absent. Macroscopic rotors may overcome this problem by exploiting in addition the rectifying effect of inertia, e.g. by some fly wheel mechanism.                 

\bigskip

{\bf Acknowledgments}
Part of this research was supported by the DFG (German Research Association) grant Sonderforschungsbereich (collaborative research center) SFB 688.

\clearpage

\appendix
\renewcommand{\theequation}{A-\arabic{equation}}
\setcounter{equation}{0}
\section*{Appendix}
In our model the motor gains free energy by running through an assembly of diffusion and chemical transition paths (Figs.~2, 4, 7, 8). 
In the absence of external forces diffusion connected cross points of diffusion and chemical transition $(IV, I); (II, III)$  exhibit the same free energy level which allows to write diffusive flow in the form of a Fick's diffusion law   Eq.~(\ref{flow1}). This was the base to derive steady state flow throughout the system, i.e. diffusive flow and chemical flux  (Eqs.~(\ref{flowmotor}, \ref{flowpotentials}). When the motor has to perform a constant workload, i.e. a constant external force is present, these cross points exhibit different free energy levels, which in the general case depend on the path the motor takes (Fig.~8). To integrate this in a more general derivation of flow one has to start with the basic Equation~(\ref{flowuni}), which relates unidirectional diffusive flow in the steady state through some domain to specific occupation number and first passage time. Unidirectional means that the domain has two boundaries, one acting as a particle source with constant concentration/probability density, the other as a pure absorber. In our model the domain consists of the diffusion paths connecting two cross points, which act as boundaries. Since no self interaction is assumed, bidirectional flow between the boundaries is just the superposition of respective unidirectional flows, i.e. 
\begin{equation}
J_{A\rightleftharpoons B}=\frac{n_{A\to B}}{\tau_{A\to B}}\rho_A - \frac{n_{B\to A}}{\tau_{B\to A}}\rho_B\;\label{Fickdomain}
\end{equation}
with $(A,B)=(IV,I),\; (II,III)$. The above Equation is a generalization of Eq.~(\ref{flow1}) and is valid in general for all type of drift forces, i.e. conservative and non-conservative, which act within the domain \cite{bauer2005}. In the steady state flow, i.e. chemical flux $J_{I\rightleftharpoons II},\; J_{III\rightleftharpoons IV}$ (Eq.~(\ref{flow3})) and diffusive flow $J_{IV\rightleftharpoons I}=J_d,\; J_{II\rightleftharpoons III}=J_p$ is constant throughout $J$ (Eq.~(\ref{flowconservation})). Hence with Eq.~(\ref{Fickdomain}) for the diffusive flows and Eqs.~(\ref{flow3}) for the chemical fluxes and Eq.~(\ref{conservationprobability}) for the conservation of probability we have a system of five linear equations determining the the four probability densities at the cross points $\rho_I\cdots\rho_{IV}$ and flow $J$. When we assume that the chemical reactions, i.e. protonation and de-protonation, are much faster than rotor diffusion, flow is obtained as  
\begin{widetext}
\begin{eqnarray}
J&=&\left[c_1\;\frac{n_{II\to III}}{\tau_{II\to III}}\;\frac{n_{IV\to I}}{\tau_{IV\to I}} - c_2\;\frac{n_{I\to
IV}}{\tau_{I\to IV}}\;\frac{n_{III\to II}}{\tau_{III\to II}}\right]\times\cr\cr\cr
&&\left[c_1\;\left(\frac{n_{II\to III}}{\tau_{II\to III}}\;  n_{IV\to
I}+\frac{n_{IV\to I}}{\tau_{IV\to I}}\; n_{II\to III}\right)+c_2\;\left(\frac{n_{III\to II}}{\tau_{III\to II}}\;  n_{{I\to IV}}+\frac{n_{I\to IV}}{\tau_{I\to IV}}\; n_{III\to II}\right)\right. \cr\cr\cr\cr 
&&\left. + \frac{k_-}{k_+} \;\left(\frac{n_{I\to IV}}{\tau_{I\to IV}}\; n_{IV\to I}+\frac{n_{IV\to I}}{\tau_{IV\to I}}\; n_{I\to
 IV}\right)+c_1c_2\;\frac{k_+}{k_-}\left(\frac{n_{II\to III}}{\tau_{II\to III}}\; n_{III\to II}+\frac{n_{III\to II}}{\tau_{III\to II}}\; n_{II\to III}\right)\right]^{-1}\label{flowgeneral}
\end{eqnarray}
\end{widetext}
The above Equation is valid for all diffusion path topologies connecting the cross points $(IV,I)$ or $(II,III)$.

\subsection*{Rotor with Check Valve Mechanism}
 We will first consider the more simple case that high barriers act as mechanical check valves and confine the rotor in the protonated state to the right, in the deprotonated state to the left path (Figs.~ 2, 3,). This arrangement avoids that the free energy becomes a multi-valued function of the diffusion path in either protonation state, a case which we will discuss later on. Hence, the free energy level $\phi$ of diffusion connected cross points differs by the workload one has to afford to get from one point to the other. With constant force $F$ on the respective diffusion paths one gets $\phi_{I}-\phi_{IV}=U_d=-F L_d$ in the deprotonated, and $\phi_{III}-\phi_{II}=U_p=-F L_p$ in the protonated state (Fig.~4). Diffusive flow vanishes when the probability densities exhibit the same chemical activity, i.e. $\rho_{IV}=e^{U_d}\rho_{I}$ and $\rho_{II}=e^{U_p}\rho_{III}$.  Together with Eq.~(\ref{Fickdomain}) this implies that
\begin{equation}
e^{-\phi_A}\frac{n_{A\to B}}{\tau_{A\to B}}=e^{-\phi_B}\frac{n_{B\to A}}{\tau_{B\to A} }\;, \label{symmetrypot}   
\end{equation}
where $(A,B)=(IV,I)$ or $(II,III)$. 
This suggests to normalize the specific occupation number by the free energy level, $\tilde n_{A\to B}=e^{-\phi_A}n_{A\to B}$. With symmetrized quantities 
\begin{eqnarray}
\tilde n&=&\frac{1} {2}\;(\tilde n_{A\to B}+\tilde n_{B\to A}) \label{symmn}\\
\tau&=&\frac{1}{2}\;(\tau_{A\to B}+\tau_{B\to A}) \label{symmnt}
\end{eqnarray}
 one obtains from Eq.~(\ref{symmetrypot}) 
\begin{equation}
\frac{\tilde n_{A\to B}}{\tau_{A\to B}}=\frac{\tilde n_{B\to A}}{\tau_{B\to A}}=\frac{\tilde n}{\tau}\;.\label{cond}
\end{equation}
This enables us to formulate a Fick's diffusion law, which is similar to that of Eq.~(\ref{flow1}), except that probability densities are replaced by corresponding activities
\begin{equation}
J_{A\rightleftharpoons B}=\frac{\tilde n}{\tau}\;(e^{\phi_A}\rho_A-e^{\phi_B}\rho_B)\; .\label{Fickm}
\end{equation}

Insertion of the normalized specific occupation numbers in Eq.~(\ref{flowgeneral}), and considering Eqs.~(\ref{cond}-\ref{Fickm}) leads to 
\begin{widetext}
\begin{eqnarray}
J&=&\frac{\frac{c_1}{c_2}e^{-U}-1}{\left(\frac{c_1}{c_2}e^{-U}+1\right)(\tau_p+\tau_d)+\left(\frac{c_1}{c_2}e^{-U}-1\right)\left(\frac{\Delta \tilde n_p}{\tilde n_p}\;\tau_p+\frac{\Delta
\tilde n_d}{\tilde n_d}\;\tau_d\right)+ 2e^{-U}c_1\; \frac{ k_+}{k_-}\frac{\tilde n_p}{\tilde n_d}\tau_d+2\frac{1}{c_2}\; \frac{k_-}{ k_+}\frac{\tilde n_d}{\tilde n_p}\tau_p}\;.
\label{flowgeneral2}
\end{eqnarray}
\end{widetext}
Here $U=U_p+U_d$ denotes the workload performed after one complete cycle, and $\Delta\tilde n=1/2(\tilde n_{A\to B}-\tilde n_{B\to A})$. One can express the above relation in terms of potentials. The driving forces are  
\begin{eqnarray}
G_i&=&-\ln\left(\frac{k_+}{k_-}\;\frac{\tilde n_p}{\tilde n_d}\;c_i\right)\cr\cr
&=&g_0 -\ln\left(\frac{\tilde n_p}{\tilde n_d}\right)-\ln(c_i)\;,\label{freeenergyappendix}
\end{eqnarray} 
where $g_0$ is the standard free energy of the protonation process. Then Eq.~(\ref{flowgeneral2}) reads
\begin{widetext}
\begin{equation}
J=\frac{1}{2}\frac{\sinh\bigg(-(\Delta G + U)/2\bigg)}{\overline{\tau}_a\cosh((\Delta G + U)/2)+\sinh(-(\Delta G + U)/2)\overline{\Delta\tau}_a+\overline{\tau}_g\cosh\left(\frac{G_1+G_2+U-\ln(\tau_d/\tau_p)}{2}\right)}\;.\label{flowgeneralappendix}
\end{equation}
\end{widetext}
Here $\overline{X}_{a}=1/2 (X_d+X_p)$  denotes the arithmetic, $\overline{X}_{g}=\sqrt{X_d X_p}$ the geometric mean of some parameter $X$ in the deprotonated and protonated state, $\Delta \tau=1/2\;(\tau_{A\to B}-\tau_{B\to A})$ is the antisymmetric counterpart to the symmetrized first passage time in Eq.~(\ref{symmnt}). In the derivation from Eq.~(\ref{flowgeneralappendix}) from Eq.~(\ref{flowgeneral2}) we exploited the relation $(\Delta \tilde n_i /\tilde n_i)\;\tau_i= \Delta\tau_i$ for each protonation state $i=d\;p$ \cite{linear}.  
\subsection*{Specific Occupation Number and First Passage Time in a Linear Potential}
A constant external force $F$ superimposes a linear potential $U(x)=-F x$ on an existing internal potential, determined by the ring neighborhood interaction. At first we will now focus solely on the first, and neglect internal interactions. We consider unidirectional steady state diffusion on a path in the protonated or deprotonated state characterized by the index $i=p,\; d$, against the external force from crosspoint $A$ at position $x=0$ to crosspoint  $B$ at position $x=L_i$. The occupation number $n_{A\to B}$ gives the probability to find the system within that path normalized by the boundary probability density $\rho_A$. From Eq.~(\ref{flow}) one can readily determine the diffusive flow in the steady state by integration when one respects the equivalence $(\partial_x-F(x))=\exp(-U(x))\partial_x\exp(U(x))$. Within a linear potential one gets  
\begin{eqnarray}
J_{A\to B}&=&D\;\bigg(\int_0^{L_i} dx\; \exp(U(x)\bigg)^{-1}\;\e^{U(0)}\;\rho_A\cr\cr\cr
&=&D\; \frac{U_i}{L_i}\;\;\frac{1}{e^{U_i}-1}\rho_A\;\label{flowappendix},
\end{eqnarray}
where $U_i=U(L_i)=-F L_i$. 
The probability density in the steady state is obtained from Eq.~(\ref{flow}) as
\begin{eqnarray}
e^{U(x)} \rho(x)-e^{U(0)}\rho_A&=&-\frac{J_{A\to B}}{D}\;\int_0^{x}d\xi\;e^{U(\xi)}\cr\cr
&=&-\frac{J_{A\to B}}{D}\;\frac{L_i}{U_i}\left(e^{U(x)}-1\right).\label{probdensityappendix}
\end{eqnarray}
Insertion of the flow of Eq.~(\ref{flowappendix}) and integration of the probability density then provides the specific occupation number as
\begin{eqnarray}
n_{A\to B}&=&\rho_A^{-1}\int_0^{L_i}dx\; \rho(x)\cr 
&=&L_i\; \left(\frac{1}{U_i}-\frac{1}{e^{U_i}-1}\right)\label{nappendix}
\end{eqnarray} 
Note that for the normalized occupation number $\tilde n_{A\to B}=n_{A\to B}$ holds, since $U(0)=0$.  
Vice versa one obtains $n_{B\to A}$ by just inverting the sign of $U_i$ in the above Equation, however for the specific occupation number one has to consider that $\tilde n_{B\to A}=e^{-U_i}n_{B\to A}$. Hence, one obtains for the symmetrized specific occupation number in either protonation state
\begin{eqnarray}
\tilde{n}_i&=&\frac{1}{2}\;(\tilde n_{A\to B}+\tilde n_{B\to A})\cr\cr
&=&\frac{1}{2}\;\frac{L_i}{U_i}\;\left(1-e^{-U_i}\right)
\end{eqnarray} 
Similarly the antisymmetric part is determined as 
\begin{eqnarray}
\Delta\tilde{n}_i&=&\frac{1}{2}\;(\tilde n_{A\to B}-\tilde n_{B\to A})\cr\cr
&=&\frac{L_i}{U_i}\;\frac{\sinh(
U_i)-U_i}{e^{U_i}-1}
\end{eqnarray} 
Up to now we only considered the influence of the external interaction on the occupation number. Potentials reflecting internal interactions have to be superimposed on $U(x)$. In the manuscript they were assumed to be constant $\Phi_i$ on the path $0<x<L_i$. Then, the occupation numbers scales with the factor 
\begin{equation}
\tilde n_i(\Phi_i)=e^{-\Phi_i}\;\tilde n_i(0)
\end{equation}
as substitution of $U(x)\to U(x)+\Phi_i$ in Eqs.~(\ref{flowappendix}-\ref{nappendix}) demonstrates. The same is true for $\Delta \tilde n$. Superimposing constant potentials on the diffusion pathways allows factorizing of the generalized equilibrium constant into a component characterizing the internal and external interaction. Therefore the corresponding potential (Eq.~\ref{freeenergyappendix}) may be written as     
\begin{eqnarray}
G_j&=&-\ln(K c_j)\cr\cr
&=&g_0-\ln\left(\frac{\tilde n_p}{\tilde n_d}\right)-\ln(c_j)\cr\cr
&=&\underbrace{g_0+\Phi_p-\Phi_d-\ln(c_j)}_{=G_{j,int}}\cr\cr
&&\underbrace{-\ln\left(\frac{1-\exp(-U_p)}{1-\exp(-U_d)}\right)}_{=G_{ext}}\;,
\end{eqnarray}
where we exploited that $U_p/L_p=U_d/L_d=-F$. 

The regular first passage time to pass an interval of length $L_i$ from one end $A$ to the other $B$ is \cite{szabo}
\begin{equation}
 \tau_{A\to B}=D^{-1}\int_0^{L_i} d\eta\;e^{U(\eta)}\;\int_0^{\eta}d\xi\; e^{-U(\xi)}\;\label{szabo}
 \end{equation}
where the diffusion coefficient $D$ was taken as constant.
For a linear potential one gets
\begin{equation}
\tau_{A\to B}=\frac{L_i^2}{D}\;\frac{e^{U_i}-U_i-1}{U_i^2}\;,
\end{equation}
and vice versa for $\tau_{B\to A}$ by changing the sign of $U_i$. Hence, the symmetrized $\tau_i=1/2(\tau_{A\to B}+\tau_{B\to A})$ and anti-symmetrized $\Delta\tau_i=1/2(\tau_{A\to B}-\tau_{B\to A})$ first passage times 
 in either protonation state are 
\begin{eqnarray}
\tau_i&=&\frac{L_i^2}{2 D}\; \frac{\sinh^2(U_i/2)}{(U_i/2)^2}\;. \cr\cr
\Delta\tau_i&=&\frac{L_i^2}{D}\;\frac{\sinh(U_i)-U_i}{U_i^2}
\end{eqnarray}
Note that the first passage time in Eq.~(\ref{szabo}) does not change when constant potentials $\Phi_i$ are superimposed, since $e^{U(\eta)+\Phi_i}\;e^{-U(\xi)-\Phi_i}=e^{U(\eta)}\;e^{-U(\xi)}$. So the symmetrized and anti-symmetrized first passage times are not affected either.  

\subsection*{Rotor and Effective Rotation - General Case}
In the case we discuss in the main paper, the free energy in the protonated state becomes a multi-valued function of the diffusion path. When we consider in Fig.~8 diffusion in the protonated state from $II$ to $III$, the motor reaches the potential 
\begin{equation}
\phi_{III}^{(r)}=U_p^{(r)}\;
\end{equation}
 when it takes the right path, but 
 \begin{equation}
 \phi_{III}^{(l)}=-U_p^{(l)}=-U_d
 \end{equation}
when it takes the left path. Hence, one cannot simply assign cross points free energy levels so that diffusive flow is driven by an activity gradient $(e^{\phi_{II}}\rho_{II}-e^{\phi_{III}}\rho_{III})$ (Eq.~\ref{Fickm}) a prerequisite to obtain flow from Eqs.~(\ref{flowgeneral2}, \ref{flowgeneralappendix}). However, one can accomplish this by introducing effective potentials $\phi^{(eff)}$.
One has to keep in mind that flow in the protonated state is the sum of flow on the left and right path
\begin{equation}
J_{II\rightleftharpoons III}=J_{p}^{(l)}+J_{p}^{(r)}\;\label{generalcase1}.
\end{equation}
Flow on each particular path is obtained from Fick's Equation (\ref{Fickm})
\begin{eqnarray}
J_{p}^{(l)}&=& \frac{{\tilde n}_p^{(l)}}{\tau_p^{(l)}}\;( \exp(\phi_{II}^{(l)})\rho_{II}-\exp(\phi_{III}^{(l)})\rho_{III})\cr\cr
&=&\frac{{\tilde n}_p^{(l)}}{\tau_p^{(l)}}\;(\rho_{II}-\exp(-U_p^{(l)})\rho_{III})\;\label{generalcase2},
\end{eqnarray}
and
\begin{eqnarray}
J_{p}^{(r)}&=& \frac{{\tilde n}_p^{(r)}}{\tau_p^{(r)}}\;( \exp(\phi_{II}^{(r)})\rho_{II}-\exp(\phi_{III}^{(r)})\rho_{III})\cr\cr
&=&\frac{{\tilde n}_p^{(r)}}{\tau_p^{(r)}}\;(\rho_{II}-\exp(U_p^{(r)})\rho_{III})\;,\label{generalcase3}
\end{eqnarray}
where we defined the cross point $(II)$ as reference point with $\phi_{II}^{(l)}=\phi_{II}^{(r)}=0$, i.e. only the free energy at point $(III)$ becomes path dependent. The parameters $\tilde n_p^{(j)}$ and $\tau_p^{(j)}$ are obtained from Eqs.~(\ref{symmnt}).
When we define  
\begin{eqnarray}
\phi_{II}^{(eff)}&=&0\cr\cr
\phi_{III}^{(eff)}&=&\ln\left[\frac{\frac{{\tilde n}_p^{(l)}}{\tau_p^{(l)}}\;\exp(\phi_{III}^{(l)})+\frac{{\tilde n}_p^{(r)}}{\tau_p^{(r)}}\;\exp(\phi_{III}^{(r)})}{\frac{{\tilde n}_p^{(l)}}{\tau_p^{(l)}}+\frac{{\tilde n}_p^{(r)}}{\tau_p^{(r)}}}\right]\cr\cr\cr
&=&\ln\left[\frac{\frac{{\tilde n}_p^{(l)}}{\tau_p^{(l)}}\;e^{-U_d}+\frac{{\tilde n}_p^{(r)}}{\tau_p^{(r)}}\;e^{U_p^{(r)}}}{\frac{{\tilde n}_p^{(l)}}{\tau_p^{(l)}}+\frac{{\tilde n}_p^{(r)}}{\tau_p^{(r)}}}\right]
\end{eqnarray}
(note $U_p^{(l)}=U_d$) we obtain with Eqs.~(\ref{generalcase1}-\ref{generalcase2}) 
\begin{eqnarray}
J_{II\rightleftharpoons III}&=&\left(\frac{{\tilde n}_p^{(l)}}{\tau_p^{(l)}}+\frac{{\tilde n}_p^{(r)}}{\tau_p^{(r)}}\right)\times\cr\cr
&&\left(\exp(\phi_{II}^{(eff)})\;\rho_{II}-\exp(\phi_{III}^{(eff)})\;\rho_{III}\right)\label{floweffpot}
\end{eqnarray}
i.e. diffusive flow is driven by a gradient of effective activities. It is worth to note that detailed balance is not fulfilled in the protonated state, since the external force superimposes a non-conservative force field. A vanishing diffusive flow in Eq.~(\ref{floweffpot}) for equal effective activities just implies that 
$J_p^{(l)}=-J_p^{(r)}$ i.e. a circular diffusive flow. This will be of importance for determination of the effective rotation. 

To bring Eq.~(\ref{floweffpot}) into the form of Eq.~(\ref{Fickm}), one has to define appropriate expression for the specific occupation number $\tilde n$ and first passage time $\tau$. According to Eq.(\ref{Fickdomain}) diffusive flow in the protonated state may written as
\begin{equation}
J_{II\rightleftharpoons III}=\frac{n_{II\to III}}{\tau_{II\to III}}\;\rho_{II}-\frac{n_{III\to II}}{\tau_{III\to II}}\;\rho_{III}\;. \label{floweffpot2}
\end{equation} 
Similarly as in Eqs.~(\ref{symmetrypot}-\ref{cond}), a vanishing flow for equal activities (Eq.~(\ref{floweffpot}) implies from Eq.~(\ref{floweffpot2}) 
\begin{equation}
\exp(-\phi_{II}^{(eff)})\frac{n_{II\to III}}{\tau_{II\to III}}=\exp(-\phi_{III}^{(eff)})\frac{n_{III\to II}}{\tau_{III\to II}}\;.
\end{equation}
 Hence, with symmetrized normalized specific occupation number and first passage time 
 \begin{eqnarray}
 \tilde n_p&=&\frac{1}{2}\;\left(\exp(-\phi_{II}^{(eff)})n_{II\to III}+\exp(-\phi_{III}^{(eff)})n_{III\to II}\right)\label{number}\cr\cr
 \tau_p&=&\frac{1}{2}\;\left(\tau_{II\to III}+\tau_{III\to II}\right)\label{rate}\;,
 \end{eqnarray}
 flow can be written finally in the form of Eq.~(\ref{Fickm})
 \begin{equation}
 J_{II\rightleftharpoons III}=\frac{\tilde n_p}{\tau_p}\;\left(\exp(\phi_{II}^{(eff)}) \rho_{II}-\exp(\phi_{III}^{(eff)})\rho_{III}\right)\;. \label{Fickm2}  
 \end{equation}
 What is left is the determination of the specific occupation numbers $n_{II\to III}\; n_{III\to II}$ and first passage times $\tau_{II\to III}\; \tau_{III\to II}$. Specific occupation numbers in the protonated state are the sum of respective numbers on each path,  $n_{II\to III}=n_{II\to III}^{(l)}+n_{II\to III}^{(r)}$, and similarly  for $n_{III\to II}$. For determination of the first passage times one exploits that flow is sum of flow on the  
left and right path, i.e. for $II\to III$
\begin{equation}
\frac{n_{II\to III}}{\tau_{II\to III}}=\frac{n_{II\to III}^{(l)}}{\tau_{II\to III}^{(l)}}+\frac{n_{II\to III}^{(r)}}{\tau_{II\to III}^{(r)}}\;,
\end{equation}
 and, hence, $1/\tau_{II\to III}$ is just the occupation number weighted average of first passage rates on the single paths $1/\tau_{II\to III}^{(l,r)}$. The same hold for parameters in direction $III\to II$. The occupation number and first passage times characterizing the left or right path are determined according to the previous section.  
 
Equation (\ref{Fickm2}) expresses diffusive flow on the assembly of left and right diffusion paths in the form of Fick's diffusion law. Hence, one can directly apply Eqs.~(\ref{flowgeneral2},  \ref{flowgeneralappendix}) to determine flow in the steady state. Note that the workload $U$ in Eq.~(\ref{flowgeneralappendix}) is replaced by an effective workload
\begin{equation}   
  U\to U^{(eff)}=U_d+(\phi_{III}^{(eff)}-\phi_{II}^{(eff)})\;.
 \end{equation}
 The efficiency factor $f$ determining the fraction of flow $J$ transformed into effective rotation, takes, respecting the constraint that the motor is confined to the left path in the deprotonated state ($J_d^{(l)}=J$),  the form 
\begin{eqnarray}
f&=&(J_d^{(l)}1-J_p^{(l)})/J \cr\cr
&=&1
-\frac{e^{\phi_{II}^{(l)}}\rho_{II}-e^{\phi_{III}^{(l)}}\rho_{III}}{e^{\phi^{(eff)}_{II}}\rho_{II}-e^{\phi^{(eff)}_{III}}\rho_{III}}
\frac{\frac{\tilde n_p^{(l)}}{\tau_p^{(l)}}}{\frac{\tilde n_p}{\tau_p}}\cr\cr
&=&1-\frac{\rho_{II}-e^{-U_d}\rho_{III}}{\rho_{II}-e^{\phi^{(eff)}_{III}}\rho_{III}}\;
\frac{\frac{\tilde n_p^{(l)}}{\tau_p^{(l)}}}{\frac{\tilde n_p}{\tau_p}}\;. 
\end{eqnarray} 
The fact that different activities act as driving forces on the left and on the left plus right path in the protonated state explains that $f$ has not the simple form as in the case without workload (Eq.~(\ref{re})). Instead one has to determine the steady state probability densities at the cross points of chemical transition and diffusion $\rho_X$, $X=I,\cdots IV$, which determines $f$ as
\begin{widetext}
\begin{eqnarray}
f&=&1-\frac{\exp(-U_d/2)}{\sinh\left(-\frac{\Delta G+U^{(eff)}}{2}\right)}\times\;\cr\cr\cr
&&\times\left[\exp\left(-\frac{\phi_{III}^{(eff)}}{2}\right)\sinh\left(-\frac{\Delta G}{2}\right) \bigg(\frac{\tilde n_p^{(l)}}{\tau_p^{(l)}}\bigg)\bigg(\frac{\tilde n_p}{\tau_p}\bigg)^{-1}+\right.\cr\cr\cr
&&\left.\exp\left(\frac{\phi_{III}^{(eff)}}{2}\right)\sinh\left(\frac{U^{(eff)}}{2}\right)\exp\left(-\frac{G_1+G_2+U^{(eff)}-\ln(\tau_d/\tau_p)}{2}\right)\right]\;,
\end{eqnarray}    
\end{widetext}
where the driving forces $G_i$ are defined in Eq.~(\ref{freeenergyappendix}), i.e. $\Delta G=-\ln(c_1/c_2)$.
Note that in the absence of external forces $(U^{(eff)}=0)$ the last summand in the bracket vanishes and $f$ takes the simple form of Eq.~(\ref{re}).

When we assume constant interaction potentials in the deprotonated  $\Phi_d$ and on the left and right path in the protonated state $\Phi_p^{(l)},\;\Phi_p^{(r)}$, and insert parameters for work against a constant force (see previous section), we obtain $f$ as a function of the workload the motor performs after one complete cycle $U=U_d+U_p^{(r)}$ (see Fig.~8), 
\begin{widetext}
\begin{eqnarray}
f&=&\frac{(e^{U}-e^{U_0})(e^{\Phi_{d}}+e^{g_0-\ln(c_1)+\Phi_p^{(l)}})}{e^{\Phi_p^{(r)}+U_p^{(l)}}e^{g_0}(c_2-c_1)\frac{e^{U_p^{(r)}}-1}{e^{U_p^{(l)}}-1}+(e^{U}-e^{U_0})(e^{\Phi_{d}}+e^{g_0-\ln(c_1)+\Phi_p^{(l)}})-e^{\Phi_d}(e^{U}-1)}\cr\cr\cr
&=&\frac{2\;\sinh\left(\frac{U-U_0}{2}\right)\;e^{\bar\Phi}}{e^{\Phi_p^{(r)}}e^{g_0}(c_2-c_1)\frac{\sinh(U_p^{(r)}/2)}{\sinh(U_d/2)}+2\;\sinh\left(\frac{U-U_0}{2}\right)\;e^{\bar\Phi}-2\;e^{\Phi_d}\;\sinh(U/2)}
\end{eqnarray}
\end{widetext}
where 
\begin{eqnarray}
\bar{\Phi}&=&\frac{1}{2}\left(\ln\left(e^{\Phi_d}+e^{\Phi_p^{(l)}+g_0-\ln c_2}\right)\right.\cr\cr
&&\left.+\ln\left(e^{\Phi_d}+e^{\Phi_p^{(l)}+g_0-\ln c_1}\right)\right)\;,
\end{eqnarray}
and  
$U_0$ is the external workload when rotation vanishes ($f=0$),
\begin{equation}
U_0=\ln\left(\frac{e^{\Phi_d}+e^{\Phi_p^{(l)}+g_0-\ln c_2}}{e^{\Phi_d}+e^{\Phi_p^{(l)}+g_0-\ln c_1}} \right)\;.
\end{equation}
Note that stabilizing the protonated state by $g_0\to -\infty$ or $\Phi_d\to +\infty$ implies $U_0\to 0$, $\bar\Phi\to \Phi_d$ and, hence, $f\to -\infty$, i.e. the rotor spins in the protonated state solely driven by the external force.  Stabilizing the right path of the protonated state $\Phi_p^{(l)}\to +\infty$ implies that the whole chemical gradient is used for directed rotation $U_0\to \ln(c_1/c_2)=-\Delta G$ and $f\to 1$.

%\end{multicols}

\begin{thebibliography}{99}
\bibitem{commentconservative}
This is due to the fact that the forces are conservative in the absence of workload.
\bibitem{linear}
This derives from the fact that when $n_{A\to B}/\tau_{A\to B}=n_{B\to A}/\tau_{B\to A}$, equal linear combination of the counters and denominators fulfill $(\alpha\;n_{A\to B}+\beta\;n_{B\to A})/(\alpha\;\tau_{A\to B}+\beta\;\tau_{B\to A})=n_{A\to B}/\tau_{A\to B}$, as long as $\alpha\tau_{A\to B}+\beta\tau_{B\to A}\neq 0$. Especially for $\alpha=\beta=1/2$ one obtains the symmetrized specific occupation number $n$ and first passage time $\tau$. Another important case is $\alpha=-\beta=1/2$, giving the antisymmetric counterpart of occupation number $\Delta n=1/2(n_{A\to B}-n_{B\to A})$ and first passage time $\Delta\tau=1/2(\tau_{A\to B}-\tau_{B\to A})$. In particular one obtains $n/\tau=\Delta n/\Delta\tau$.
\end{thebibliography}
\end{document}